\documentclass[11pt]{article}
\usepackage{subfigure}
\usepackage{graphicx}
\usepackage{cancel}
\usepackage{cite}
\usepackage{cleveref}
\usepackage{lipsum}
\usepackage{amsmath}
\usepackage{indentfirst}
\usepackage{multirow}
\usepackage{amssymb}
\usepackage{feynmp}
\usepackage[utf8]{inputenc}
\usepackage{hyperref}

\usepackage{amsmath,amssymb,amsthm,amsxtra,overpic,bbm,bm,epsfig,ulem}
\usepackage{booktabs}
\usepackage{mathrsfs}
\usepackage{graphicx}% Include figure files
\usepackage{dcolumn}% Align table columns on decimal point
\usepackage{bm}% bold math
\usepackage{amsmath}
\usepackage{slashed}  % slashed p
\usepackage{setspace}
\usepackage{color}

\begin{document}
\unitlength=1mm
\begin{center}
	{\Large \bf Single production of vector-like bottom quark at the LHeC}
\end{center}

\vspace{0.5cm}

\begin{center}
{\bf Xue Gong \footnote{E-mail: gongxue422@163.com},  Chong-Xing Yue \footnote{E-mail: cxyue@lnnu.edu.cn}, Hai-Mei Yu \footnote{E-mail: yuhaimei233@163.com} and Dong Li\footnote{E-mail: lidonglnnu@163.com}} \\
{Department of Physics, Liaoning Normal University, Dalian 116029, China}
\end{center}

\vspace{4cm}
\begin{abstract}

Existences of vector-like quarks (VLQs) are predicted  in many new physics scenarios beyond the Standard Model (SM).
We study the possibility of detecting the vector-like bottom quark (VLQ-$B$) being the $SU(2)$ singlet with electric charge $-1/3$ at the Large Hadron Electron Collider (LHeC) in a model-independent framework. The decay properties and single production of VLQ-$B$ at the LHeC are explored. Three types of signatures are investigated.
By carrying out a fast simulation for the signals and the corresponding backgrounds, the signal significances are obtained.
Our numerical results show that detecting of VLQ-$B$ via the semileptonic channel is better than via the fully hadronic or leptonic channel.

\vspace{8cm}	
\end{abstract}

\vspace{18cm}

\section*{\uppercase\expandafter{\romannumeral1}. Introduction}

With the discovery of a 125~GeV Higgs boson in July 2012 by the ATLAS and CMS collaborations at the CERN Large Hadron Collider (LHC)\cite{Aad:2012tfa,Chatrchyan:2012xdj},
the Standard Model (SM) has acquired remarkably success at explaining most of the available experimental phenomena with great accuracy.
As yet, there are still unresolved theoretical issues in the SM, such as the nature of the electroweak symmetry breaking and the hierarchy between the electroweak and the Planck  scales.
~One solution is by introducing new heavy particles called vector-like quarks (VLQs)
which regulate the Higgs boson mass-squared divergence\cite{DeSimone:2012fs,Aguilar-Saavedra:2013qpa}.%1802.01486
~Since VLQs can obtain the gauge invariant mass terms of  the form $m\bar{\psi}\psi$ directly, they are not subject to the constraints from Higgs production.
Therefore, VLQs as a class of interesting particles have not been excluded by precision measurements.

~VLQs are hypothetical spin-1/2 colored fermions and are proposed in many new physics scenarios, for example, little Higgs\cite{ArkaniHamed:2002qy,ArkaniHamed:2002qx,
Schmaltz:2002wx,Schmaltz:2005ky}, composite Higgs\cite{Kaplan:1983sm,Agashe:2004rs,Anastasiou:2009rv,Low:2015nqa} and extra dimensions models\cite{Randall:1999ee,Cheng:1999bg}.
The left- and right-handed components of VLQs have the same transformation properties under the SM electroweak symmetry group\cite{delAguila:1982fs,AguilarSaavedra:2009es}.
%A common feature of these new fermions is that they are assumed to decay to a SM quark with a SM gauge boson, or a Higgs boson.
%
	VLQs can be embedded in singlet [$T$, $ B$], doublets [ $\left(X,T\right),\left(T,B\right)$ or $\left(B,Y\right)$] or triplets [$\left(X,T,B\right)$ or $\left(T,B,Y\right)$] for the representations of the $SU(2)$ group.
	The weak hypercharges of VLQs can be determined by their Yukawa couplings with the SM quarks and the Higgs boson,
	therefore there are four possible charge assignments: $Q_T= +2/3$, $Q_B= -1/3$, $Q_X= +5/3$, and $Q_Y= -4/3$.
	In the case of UV complete models, the extra-dimensional models predict a tower of the vector-like top quarks, of which the lightest one has sizable mixing with the third generation quarks~\cite{delAguila:2000kb}.
	The vector-like  bottom quarks  are predicted in grand unification theories based on $E^{}_6$~\cite{Frampton:1999xi,Hewett:1985}.
	The doublets $\left(X,T\right)$ and $\left(T,B\right)$  naturally emerge in warped models~\cite{Contino:2006qr,Carena:2006bn}.
	The triplets $\left(X,T,B\right)$ and $\left(T,B,Y\right)$ are predicted in the composite Higgs models~\cite{Aguilar-Saavedra:2019ghg}.
In this paper, we focus on the $SU(2)$ singlet vector-like bottom quark (VLQ-$B$) in a model-independent way.

A lot of phenomenological studies for VLQs have been presented in vast literatures~\cite{Fuks:2016ftf}.
Ref.\cite{Vignaroli:2012sf} has considered single production of VLQ-$B$ which decays into $Hb$ at the LHC in context of the composite Higgs model.
Ref.\cite{Nutter:2012an} introduced an effective Lagrangian to study the possibility of detecting VLQ-$B$ via the decay channel $B \to Wt$ at the LHC.
Ref.\cite{Cheung:2020vqm} performed global fits of the constraints of VLQ-$B$ by using the CKM unitarity violation, excess in Higgs signal strength, and bottom quark forward-backward asymmetry.
In our previous work\cite{Gong:2019zws}, we have considered the capability of detecting VLQ-$B$ at the LHC via single production channel, which is a more potential process than pair production since its less phase-space suppression when the mass of VLQ-$B$ is more heavier.

The concrete search strategies depend on the decay modes of VLQs. If only coupling to the third-generation quarks, the vector-like top quark (VLQ-$T$) and VLQ-$B$ could decay via the charged current, i.e. $T \to Wb$ and $B \to Wt$, or neutral current, i.e. $T \to Ht, Zt$ and $B \to  Hb, Zb$. The subsequent decays of the SM particles produce rich signals. In order to distinguish the type of VLQs, discriminant analysis is required.
For example, in the case of fully hadronic final state which is powerful for the $B\to Hb$ channel, Ref.~\cite{Aaboud:2018wxv} introduced a discriminant function $P$ that can tag the parent particles of the jets as $W/Z$ boson, Higgs boson or $t$ quark. Together with $b$-tagging and the invariant mass of jets, it is convenient to reduce the background and distinguish signal events of the VLQ-$T$ and VLQ-$B$.
Ref.~\cite{Aaboud:2018pii} combines the searches for pair-produced VLQ-$T$ and VLQ-$B$ at the LHC with all three possible decay modes,
the lower limit of singlet VLQ-$B$ mass is a bit larger than that of singlet VLQ-$T$ mass, and the singlet VLQ-$B$ is more sensitivity to the channel $B \to Wt$,
while the singlet VLQ-$T$ is more sensitivity to the channel $T \to Ht$.

By now, the direct searches for VLQ-$B$ have been performed by the ATLAS and CMS collaborations at the LHC with center-of-mass (c.m.) energy of $\sqrt{s}$ = 13 TeV  and an integrated luminosity of 35-36 fb$^{-1}$\cite{Sirunyan:2018omb,Aaboud:2018uek,Sirunyan:2018fjh}. Although there are not any signatures be detected, the constraints on VLQ-$B$ have been obtained. The most stringent bounds on the VLQ-$B$ mass are in the range of 700–1800~GeV depending on the production modes, the considered final states and the assumed branching ratios.
In fact, the collider has become and will remain an important tool to test wide classes of new physics models.
Thus, it is highly motivated to investigate all sensitive search strategies within the possibly available accelerator and detector designs.

Here, we intend to study the possibility of detecting VLQ-$B$ in the proposed powerful high energy $ep$ collider, the Large Hadron Electron Collider (LHeC)\cite{AbelleiraFernandez:2012cc}  with a 60-140 GeV electron beam and a 7 TeV proton beam from the LHC.
It is supposed to run synchronously with the HL-LHC and to deliver the integrated luminosity of 100 fb$^{-1}$ per year and of 1000 fb$^{-1}$ in total.
Compared to the previous $ep$  ollider, HERA, the LHeC extends one order of magnitude in the c.m. energy and 1000 times in the integrated luminosity.
Refs.\cite{Liu:2017rjw,Zhang:2017nsn,Han:2017cvu} have studied the discovery potential of the VLQ-$T$ through various channels at the LHeC, where the VLQ-$T$ is the $SU(2)$ singlet with charge 2/3.
To the best of our knowledge, so far, no work has been done to search single production of the $SU(2)$ singlet VLQ-$B$ at the LHeC.
Hence, we mainly study the observability of the single VLQ-$B$ production at the LHeC combine with the $B \to Wt$ decay channel in our work.
Considering the final state has two $W$-bosons (one of those come from top quark decaying), we analyze three types of signatures, which come from the fully hadronic decay channel, the fully leptonic decay channel, and the semileptonic decay channel, respectively. We expect that such work  may become a complementary to other production processes in searches for the heavy VLQ-$B$ at the LHeC.

This paper is organized as follows: in section II, we brief review  the couplings of VLQ-$B$ with the SM particles, and discuss its possible decay modes.
Section III devotes to a detailed analysis of the relevant signals and backgrounds at the LHeC. Finally, we summarize our results in section IV.

\section*{II. Effective Lagrangian and decay modes of the vector-like bottom quark}

VLQs can interact with the SM quarks and the Higgs boson through Yukawa couplings. After the Higgs developing a nonzero vacuum expectation value (VEV),  VLQs are allowed to mix with the SM quarks. The mass matrices of quarks are determined by the chosen $SU(2)$ representations of VLQs.
By diagonalizing the mass matrices, one can obtain the couplings between physical states which can be found in Ref.\cite{Okada:2012gy}.
Ref.\cite{Buchkremer:2013bha} proposed a more compact parameterization for VLQ-$T$ couplings.
Similarly, we consider the same parameterization in the case of VLQ-$B$ and assume that it is  the $SU(2)$ singlet.
The generic parametrization of an effective Lagrangian of VLQ-$B$ is given by (showing only the couplings relevant for our analysis):
\begin{eqnarray}
\mathcal{L} &=& \dfrac{\kappa^{}_B}{2}
\Biggr \{
\sqrt{ \dfrac{ R^{}_L }{1 + R^{}_L} } \dfrac{g}{\sqrt{2}} ~[ \bar{B}^{}_L W^-_\mu \gamma^\mu u^{}_{L}]
+ \sqrt{ \dfrac{1}{1+R^{}_L}} \dfrac{g}{ \sqrt{2}} ~[ \bar{B}^{}_L W^-_\mu \gamma^\mu t^{}_L ]  \nonumber\\
&&+ \sqrt{ \dfrac{R^{}_L}{1+R^{}_L} } \dfrac{g}{2C^{}_W} ~[\bar{B}^{}_L Z^{}_\mu \gamma^\mu d^{}_{L} ]
+ \sqrt{ \dfrac{1}{1+R^{}_L}} \dfrac{g}{2C^{}_W} ~[ \bar{B}^{}_L Z^{}_\mu \gamma^\mu b^{}_L ] \nonumber\\
&&- \sqrt{ \dfrac{R^{}_L}{1+R^{}_L}} \dfrac{M^{}_B}{\upsilon} ~[ \bar{B}^{}_R H d^{}_{L} ]
- \sqrt{ \dfrac{1}{1+R^{}_L}} \dfrac{M^{}_B}{\upsilon} ~[ \bar{B}^{}_R H b^{}_L] - \sqrt{ \dfrac{1}{1+R_L}} \dfrac{m^{}_b}{\upsilon} ~[ \bar{B}^{}_L H b^{}_R]
\Biggr \} + {\rm h.c.} \;,
\label{Lag1}
\end{eqnarray}
where $u_{}(d_{})$ denotes the first or second generation  up(down)-type quark, $g$ is the $SU(2)$  coupling constant, $\upsilon \simeq 246$ GeV is the electroweak symmetry breaking scale.
We have abbreviated $\cos\theta^{}_W$ as $C^{}_W$, where $\theta^{}_W$ is the Weinberg angle.
~There are only three parameters that fully describe the relevant interactions we consider.
~Besides the mass parameter $M^{}_B$, there are two coupling parameters appearing in Eq.(\ref{Lag1}):
\begin{itemize}
\item $\kappa^{}_B$, the coupling strength to SM quarks in units of standard couplings, which is only relevant to single production;
\item $R^{}_L$, the generation mixing coupling, which describes the rate of decays to first two generation quarks with respect to the third generation, where the subscript $L$ represents the chirality of the fermions. For the singlet VLQ-$B$, we neglect the mixing of right-handed quarks since it is suppressed \cite{Aguilar-Saavedra:2013qpa}.
\end{itemize}
From Eq.(\ref{Lag1}) we can see that, for the couplings of the singlet VLQ-$B$ with the SM fermions, $R^{}_L = 0$ corresponds to coupling to top and bottom quarks only.
In many new physics models, VLQs are expected to couple preferentially to third-generation quarks~\cite{AguilarSaavedra:2009es,delAguila:1982fs} 	and can have flavor-changing neutral-current decays in addition to charged-current decays~\cite{Aaboud:2018pii}.
While the limit $R^{}_L = \infty$ represents coupling to first and second generation of quarks only,	
which are not excluded~\cite{Aguilar-Saavedra:2013qpa,Atre:2008iu} although not favored.
  \begin{figure}[htbp]
	\centering
	\includegraphics[scale=0.78]{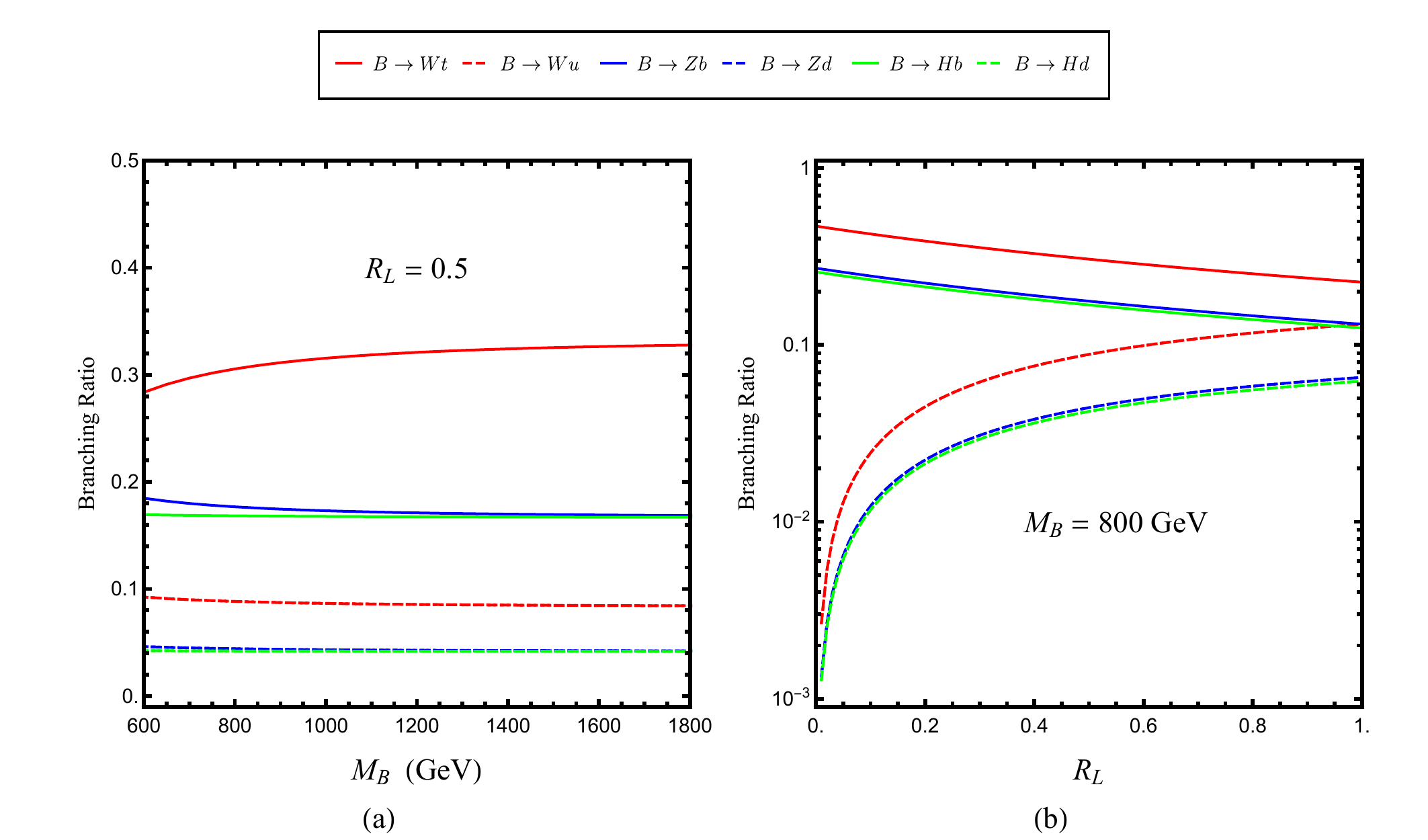}
	\caption{The branching ratios of the decay modes  $Wu^{}_i$, $Zd^{}_i$ and $Hd^{}_i$ as functions of the \hspace*{1.65cm} VLQ-$B$ mass $M^{}_B$ (left) and the generation mixing coupling $R^{}_L$ (right).}
	\label{fig1}
\end{figure}

According to above discussions, VLQ-$B$ has three typical of decay modes: $Wu^{}_i$, $Zd^{}_i$, and $Hd^{}_i$, where $i = 1, 2, 3$.
The corresponding partial widths are given by
 \begin{eqnarray}
&&\hspace{-0.8cm} \Gamma \left( B \to Wu^{}_i \right) = \frac{c^{}_i e^2 \kappa_B^2 M_B^3 } {256 \pi S_W^2 m_W^2 } \lambda^{1/2} \left( 1 , \frac{m_{u^{}_i}^2}{M^2_B} , \frac{m_W^2}{M^2_B} \right)
 \Big[ \left( 1 - \frac{m_{u^{}_i}^2}{M_B^{}} \right)^2 + \frac{m_{u^{}_i}^2 m_W^2}{M^4_B} + \frac{m_W^2}{M^2_B} - 2 \frac{m_W^4}{M^4_B} \Big] \;,
 \\
&&\hspace{-0.8cm} \Gamma \left( B \to Zd^{}_i \right) = \frac{c^{}_i e^2 \kappa_B^2 M_B^3 } {512 \pi S_W^2 m_W^2 } \lambda^{1/2} \left( 1 , \frac{m_{d^{}_i}^2}{M^2_B} , \frac{m_Z^2}{M^2_B} \right)
 \Big[ \left( 1 - \frac{m_{d^{}_i}^2}{M_B^{}} \right)^2 + \frac{m_{d^{}_i}^2 m_Z^2}{M^4_B} + \frac{m_Z^2}{M^2_B} - 2 \frac{m_Z^4}{M^4_B} \Big] \;,
 \\
&&\hspace{-0.8cm} \Gamma \left( B \to Hd^{}_i \right) = \frac{c^{}_i \kappa_B^2 M_B^{} } {128 \pi } \lambda^{1/2} \left( 1 , \frac{m_{d^{}_i}^2}{M^2_B} , \frac{m_h^2}{M^2_B} \right)
 \Big[ \left( g_p^2 + g_m^2 \right) \left( 1 + \frac{m_{d^{}_i}^2}{M^2_B} - \frac{m_h^2}{M^2_B} \right) + 4 g_m^{} g_p^{} \frac{m_{d^{}_i}^{}}{M^{}_B} \Big] \;,
 \end{eqnarray}
where $c^{}_i = 2/\left(1+R^{}_L\right)$ for $t,b$ quarks, and $c^{}_i =  R^{}_L/\left(1+R^{}_L\right)$ for $u,c,d,s$ quarks,
 \begin{eqnarray}
 g^{}_m = \frac{M^{}_B}{v} \;,\quad g^{}_p = \frac{m^{}_b}{v} \;,\quad
 \end{eqnarray}
 and the function $\lambda(a,b,c)$ is given by
 \begin{eqnarray}
  \lambda\left(a,b,c\right) = a^2 + b^2 + c^2 - 2 ab - 2 ac - 2 bc \;.
 \end{eqnarray}
 In the limit of $M^{}_B \gg m^{}_t$, the partial widths can be approximate written as
 \begin{eqnarray}
 \Gamma \left( B \to Wu^{}_i \right) &\simeq& \frac{c^{}_i e^2 \kappa_B^2 M_B^3 } {256 \pi S_W^2 m_W^2 } \;,
 \label{Lag7}
 \\
 \Gamma \left( B \to Zd^{}_i \right) &\simeq& \frac{c^{}_i e^2 \kappa_B^2 M_B^3 } {512 \pi S_W^2 m_W^2 } \;,
  \label{Lag8}
 \\
  \Gamma \left( B \to Hd^{}_i \right) &\simeq& \frac{c^{}_i \kappa_B^2 M_B^3 } {128 \pi v^2} = \frac{c^{}_i e^2 \kappa_B^2 M_B^3 } {512 \pi S_W^2 m_W^2 } \;.
   \label{Lag9}
 \end{eqnarray}
From above equations, we can see that, for heavy VLQ-$B$,
$\dfrac{1}{2} \Gamma \left( B \to Wu^{}_i \right) \simeq \Gamma \left( B \to Zd^{}_i \right) \simeq \Gamma \left( B \to Hd^{}_i \right)$ is a good approximation as expected by the Goldstone boson equivalence theorem\cite{Lee:1977eg}.
From the Lagrangian given in Eq.(\ref{Lag1}), one may expect the branching ratios of $Hd^{}_i$ channel is the largest one since the coupling coefficient of $BHd^{}_i$ is proportional to $M^{}_B$.
Actually, the partial width $ \Gamma \left( B \to Hd^{}_i \right) $ is  proportional to  $M^{3}_B$, which is similarly with that for other two decay channels as shown in Eqs.(\ref{Lag7} -\ref{Lag9}).
The branching ratios of these decay channels are plotted as functions of the mass parameter $M_B$ and the generation mixing coupling $R^{}_L$ in Fig.~\ref{fig1}. Since the mass of the boson $W$, $Z$ or $H$ is much larger than that of  the first or second generation quark, the branching ratios of  VLQ-$B$ decaying to the first and second generation quarks are approximately equal each other, so we only give the branching ratios of  VLQ-$B$ decaying to the first generation quarks in Fig.~\ref{fig1}.
For $M_B \geq 800 $ GeV, one can see that the branching ratios approximate satisfy
$Br \left(B\rightarrow Wu^{}_i\right): Br \left(B\rightarrow Zd^{}_i\right): Br \left(B\rightarrow Hd^{}_i\right) \approx 2:1:1$.
As expected, the branching ratios of the first and second generation quarks vanish rapidly when $R^{}_L$ approaches to zero.
When $R^{}_L <1$, the third generation quarks plus bosons are the main decay channels.
Hence,  we choose the $Wt$ channel to study the possibility of detecting the signals of VLQ-$B$ at the LHeC in our work.

\section*{III. Signal analysis and discovery potentiality}

\begin{figure}[htbp]
\centering
\includegraphics[height=5cm, width=8cm]{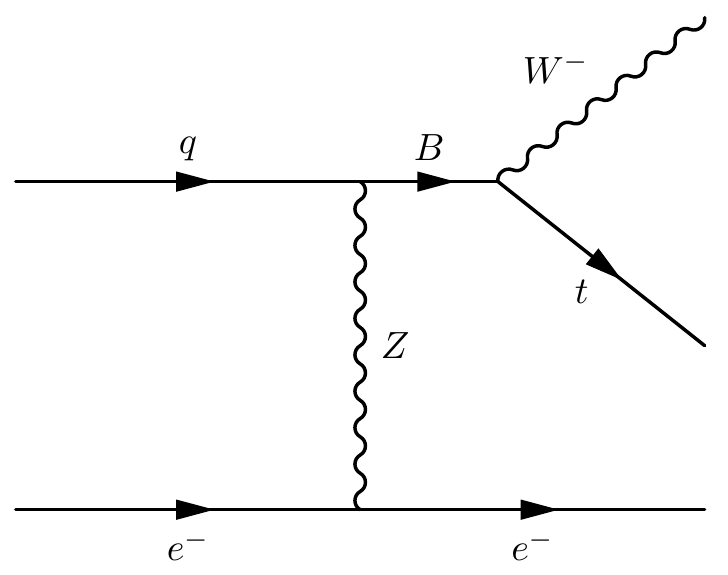}
\includegraphics[height=5cm, width=8cm]{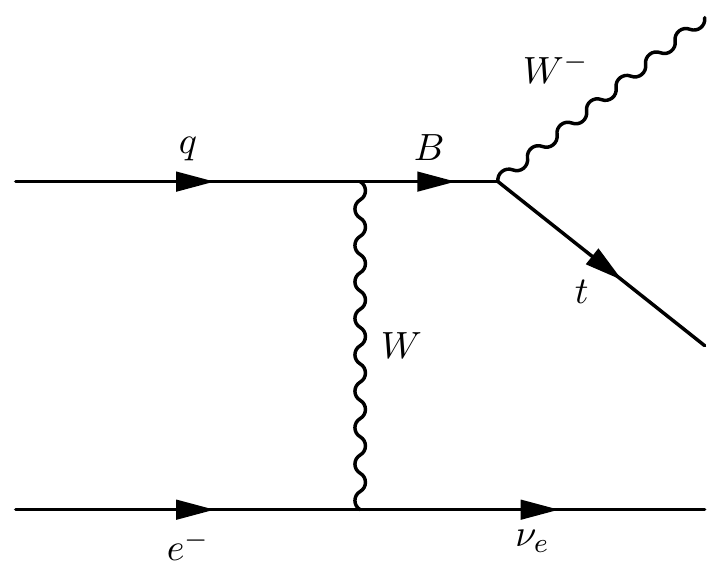}
\caption{The Feynman diagrams for single production of VLQ-$B$  at the LHeC including the decay \hspace*{1.75cm}channel $B \rightarrow Wt$.}
\label{fig2}
\end{figure}

For the single production of VLQ-$B$ at the LHeC, the dominant way is mediated by the exchange of a $W$ or $Z$ boson in the $t$-channel,
the $H$-mediated process can be ignored due to the tiny Yukawa coupling between electron and Higgs boson.
The relevant Feynman diagrams for the single production and decaying into $Wt$ are presented in Fig.~\ref{fig2}.
For the chosen decay channel of VLQ-$B$, the final state contains two $W$-bosons (one of those coming from top quark decay).
There are three types of signatures, which come from the fully hadronic, the fully leptonic and the semileptonic decay channels, respectively.
\begin{figure}[htbp]
	\centering
	\includegraphics[scale=0.8]{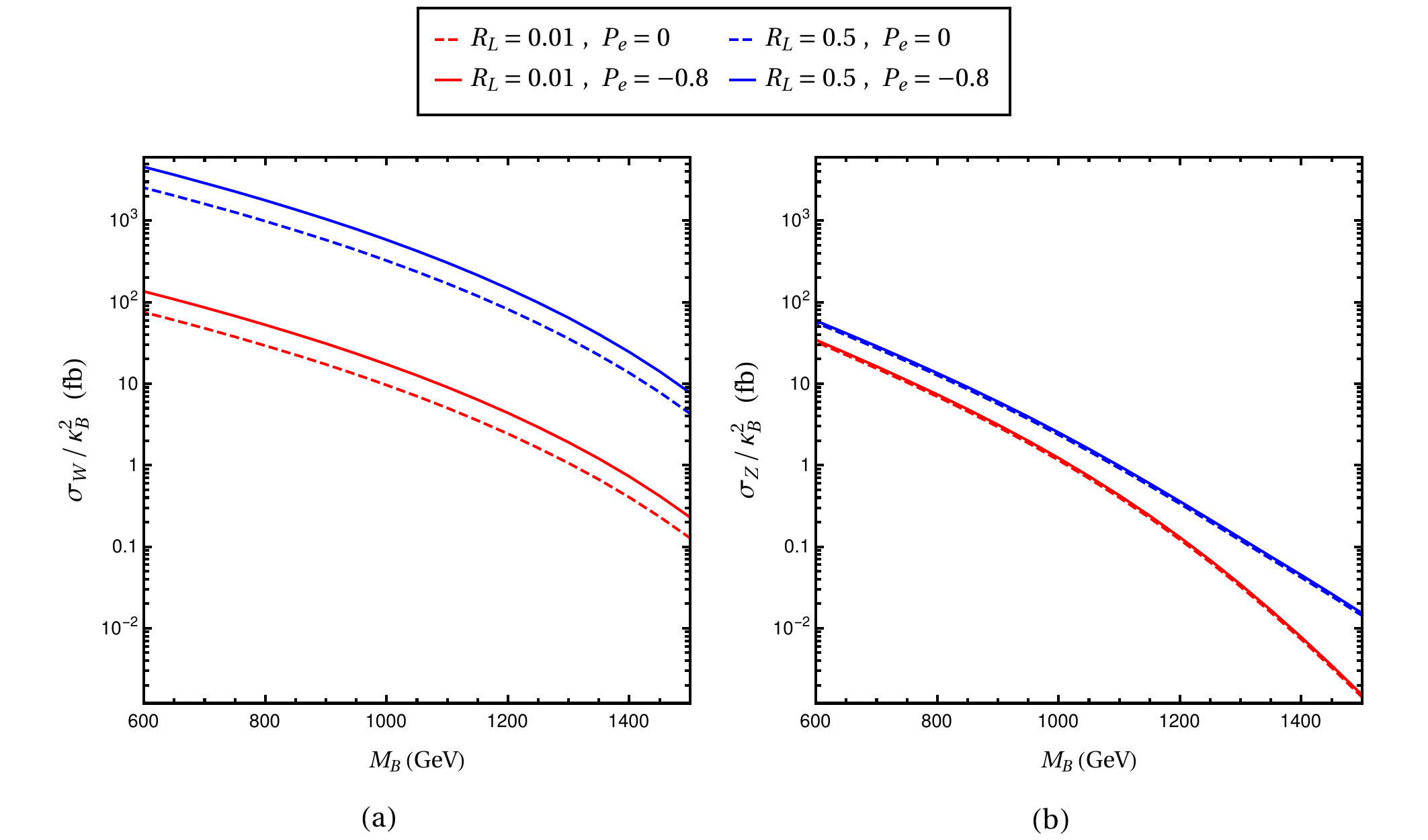}
	\caption{The scaled cross sections of the processes $e^- p \to W \to \nu B$ and  $e^- p \to Z \to e^- B$ as \hspace*{1.65cm} functions of $M^{}_B$ for
	 different $R^{}_L$ and polarization of $e^-$ beam.}
	\label{figXSMB}
\end{figure}

\begin{figure}[htbp]
	\centering
	\includegraphics[scale=0.8]{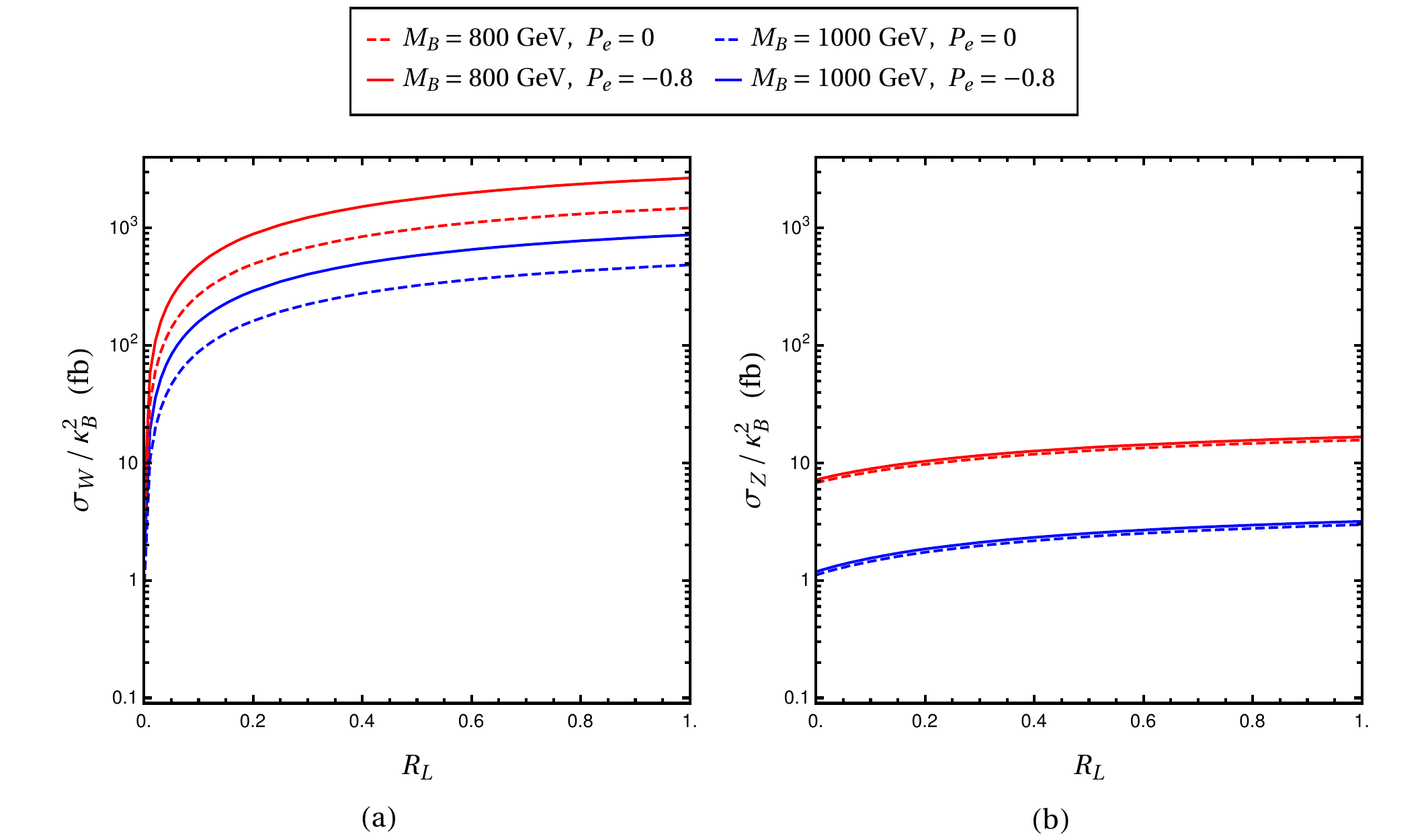}
	\caption{The scaled cross sections of the processes   $e^- p \to W \to \nu B$ and  $e^- p \to Z \to e^- B$ as  \hspace*{1.65cm} functions of $R^{}_L$ for
		different $M^{}_B$ and polarization of $e^-$ beam.}
	\label{figXSRL}
\end{figure}
To proceed further analysis, we need to know the values of some parameters. The SM input parameters which relevant to our calculations are taken from Ref.\cite{Tanabashi:2018oca} as follow:
\begin{eqnarray}
&& m^{}_t = 173.0 ~\mathrm{GeV} \;,\quad m^{}_Z = 91.1876~\mathrm{GeV} \;,\quad m^{}_h = 125~\mathrm{GeV}, \nonumber \\
&& S^2_W = 0.231 \;,\quad \alpha\left(m^{}_Z\right) = 1/128.
\end{eqnarray}
Considering the current constraints on the VLQ-$B$ mass\cite{Sirunyan:2018omb,Aaboud:2018uek,Sirunyan:2018fjh},
we choose three benchmark points: $M_B = 800, 900, 1000 $ GeV, which are referred to as $B_{800}$, $B_{900}$, $B_{1000}$, respectively. The stringent bounds on the coupling parameter $\kappa_B$ come from the experimental data about the $Z\bar bb$ couplings, which give the upper limit as  $\kappa^{}_B < 0.23$~\cite{Buchkremer:2013bha}.
%In our numerical estimation, we will take $\kappa^{}_B =0.1$.
For taking the $e^-$ and $p$ beam energy as 140~GeV and 7~TeV respectively, the c.m. energy of the LHeC is $\sqrt{s} = 1.98$ TeV.
In Fig. \ref{figXSMB} and \ref{figXSRL} , we show the scaled cross sections of the processes  $e^- p \to Z \to e^- B$ and $e^- p \to W \to \nu B$ with different $M^{}_B$, $R^{}_L$ and $e^-$ beam polarization at the LHeC.
From these figures, one can see that: (1) The cross section of $W$-mediated process can be enhanced by the polarization of $e^-$ beam, while that of $Z$-mediated process is insensitive, since $W$ boson only couple to the left-handed electron.
(2) The cross section of $W$-mediated process is more sensitive to $R^{}_L$ than that of $Z$-mediated process. And the cross section of $W$-mediated process vanishes when $R^{}_L$ approaches to zero, the reason is that no top parton in proton. 
~Hence, in the following, we choose $W$-mediated process for simulation in the case of $ P_{e}=-0.8$, $\kappa^{}_B = 0.1$, and $R^{}_L=0.5$. 

The model file\cite{UFO} which realize the Lagrangian given by Eq.(\ref{Lag1}) can be found in the dedicated FeynRules package\cite{fr}.
Signal and background events are simulated at the leading order using MadGraph5-aMC@NLO\cite{mg} with the CTEQ6L parton distribution function (PDF) \cite{Pumplin:2005rh}.
Showering, fragmentation and hadronization are performed with customized Pythia~\cite{pythia}.
The PGS is applied for detector simulation, and the relevant parameters are taken for LHeC Detector Design~\cite{AbelleiraFernandez:2012cc, 22}.
The anti-$\kappa_t$ algorithm \cite{Cacciari:2008gp}  with parameter $\Delta R = 0.4$ is used to reconstruct the jets.
Finally, MadAnalysis5~\cite{md} is applied for data analysis and plotting.
The energy resolutions of lepton and jets are taken as~\cite{AbelleiraFernandez:2012cc}
\begin{eqnarray}
\frac{\Delta E}{E} = \frac{a}{\sqrt E} \oplus  b \;,
\end{eqnarray}
where $a= 0.45~{\rm GeV}^{-1}$ , $b=0.03$ for jets and $a= 0.085~{\rm GeV}^{-1}$ , $b=0.003$ for leptons.
Assuming the efficiencies of LHeC detector are the same with those of CMS detector,
then the efficiency of muon is set as $0.95$, the efficiency of electron is set as $0.95$ for $\left|\eta\right| \leq 1.5$ and $0.85$ for $1.5\leq \left|\eta\right| \leq 2.5$.
The efficiency for tagging $b$-quark jets is $85\%$, the light-parton misidentification probability is $10\%$~\cite{Chatrchyan:2012jua}.

\subsection*{A. The fully hadronic channel }

In this subsection, we analyze the signal and background events and explore the sensitivity of the singlet VLQ-$B$ at the LHeC ($\sqrt{s} = 1.98$ TeV) through the fully hadronic decay channel:
\begin{equation}
e^- p\to \nu B(\to W^-t) \to \nu W^-(\to j j)t(\to W^+b\to j j b)\to  4j+b + \cancel{E}^{}_T.
\end{equation}
For this channel, the typical signal is one $b$-jet, four jets and large missing energy. The main SM backgrounds come from the following five processes:
\begin{itemize}
	\item  BKG1: $\nu \bar t+$jets:	$\nu  p \to \nu \bar t + {~\rm jets} \to \nu  W^- \bar b + {~\rm jets} \to \bar b+\cancel{E}^{}_T + {~\rm jets} $,
	\item  BKG2: $\bar \nu V+$jets:	$e^- p \to \nu V + {~\rm jets} \to \cancel{E}^{}_T + {~\rm jets} $,
	\item  BKG3: $\nu t \bar t+$jets:	$\nu  p \to \nu t \bar t + {~\rm jets} \to \nu  W^- \bar b W^+ b + {~\rm jets} \to b\bar b+\cancel{E}^{}_T + {~\rm jets} $,	
    \item  BKG4: $\bar \nu V V+$jets:	$e^- p \to \nu V V + {~\rm jets} \to \cancel{E}^{}_T + {~\rm jets} $,
    \item  BKG5: $\bar \nu V \bar t+$jets:	$e^- p \to \nu V \bar{t} + {\rm ~jets} \to \nu V W^- \bar b + {~\rm jets} \to \bar b+\cancel{E}^{}_T + {~\rm jets} $,
\end{itemize}
where $V$ denotes $W$ or $Z$ boson, the one light jet mentioned above can be faked as $b$-jet.
 To avoid double counting for multiple jets and parton shower, the MLM~\cite{Hoche:2006ph} matching method with xqcut = 25 GeV is applied.
The signal and background processes are simulated at the  LHeC with the integrated luminosity of 1000 fb$^{-1}$. Firstly, we apply the basic cuts to the signal and background events, which are used to simulate the geometrical acceptance and detection threshold of the detector. These basic cuts are selected as follows in our simulation
\begin{eqnarray}
 p^{j}_{T}> 20~{\rm GeV} \;,\quad \vert\eta^{j}\vert < 5 \;,\quad \Delta R(x,y) > 0.4 \;, \nonumber
\end{eqnarray}
where the particle separation $\Delta R_{xy}$ is defined as $\sqrt{ (\Delta \eta^{}_{xy})^2 + (\Delta \phi^{}_{xy})^2 }$ with $\Delta \eta^{}_{xy}$ and $\Delta \phi^{}_{xy}$ being the rapidity and azimuthal angle gaps between the two particles $x$ and $y$.

\begin{figure}[http]
	\subfigure[]{
		\begin{minipage}[t]{0.5\textwidth}
			\includegraphics[height=6cm, width=11cm]{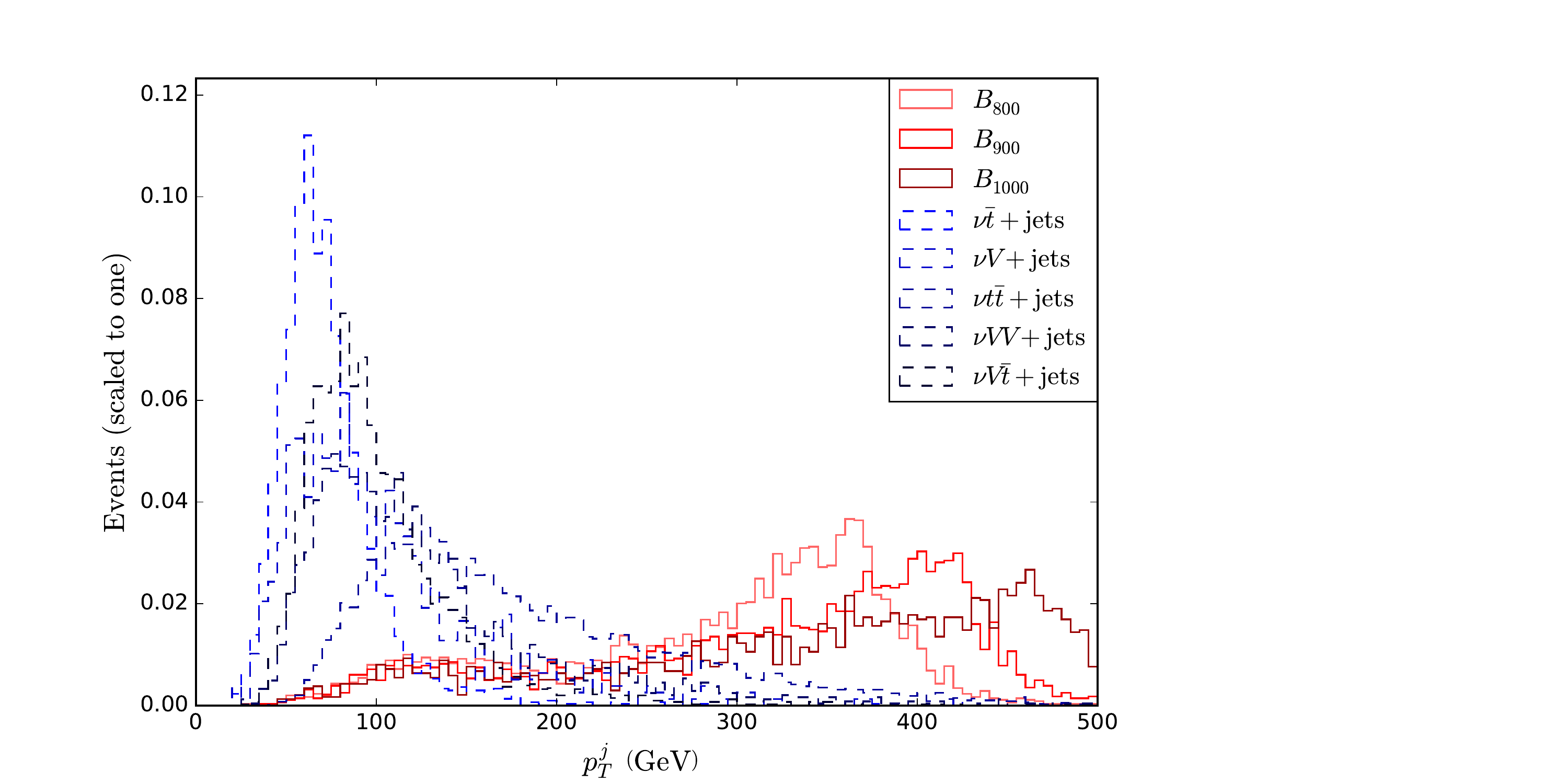}
		\end{minipage}
		\label{fig4a}}
	\subfigure[]{
				\begin{minipage}[t]{0.5\textwidth}
		\includegraphics[height=6cm, width=11cm]{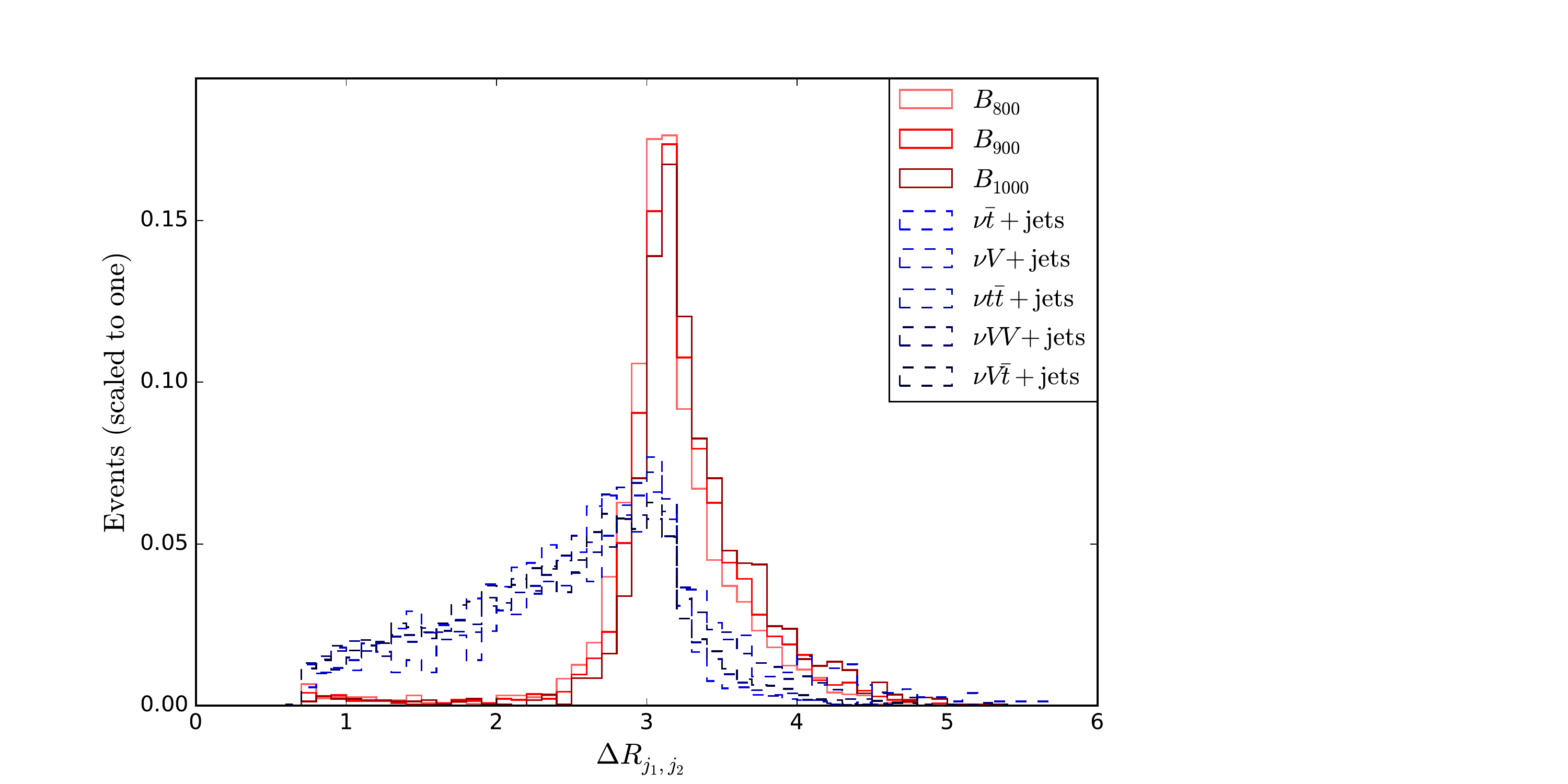}
				\end{minipage}
		\label{fig4b}
	}		
	
	\subfigure[]{
		\begin{minipage}[t]{0.5\textwidth}	
			\includegraphics[height=6cm, width=11cm]{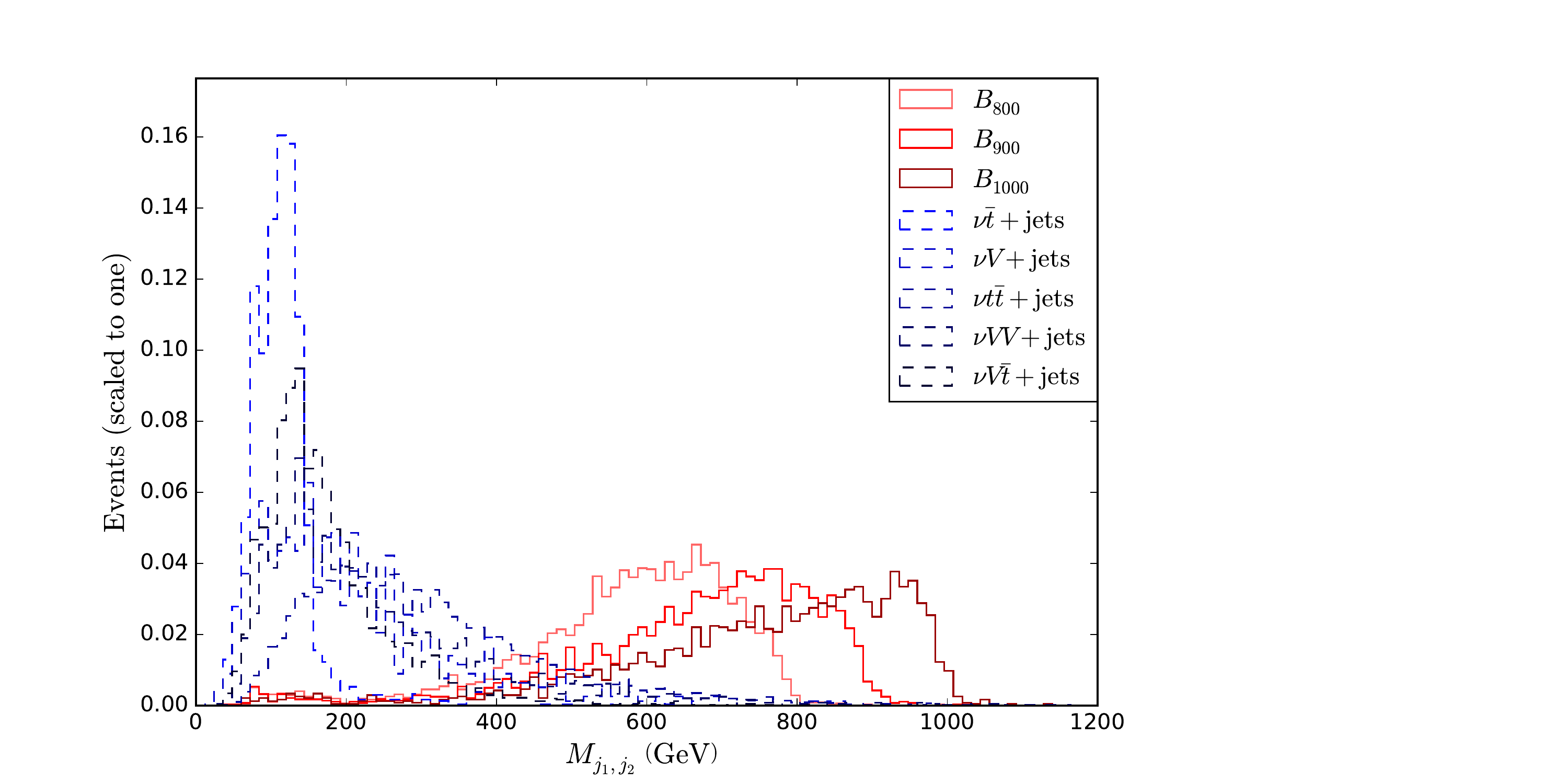}	
		\end{minipage}	
		\label{fig4c}
	}
	\subfigure[]{
				\begin{minipage}[t]{0.5\textwidth}
		\includegraphics[height=6cm, width=11cm]{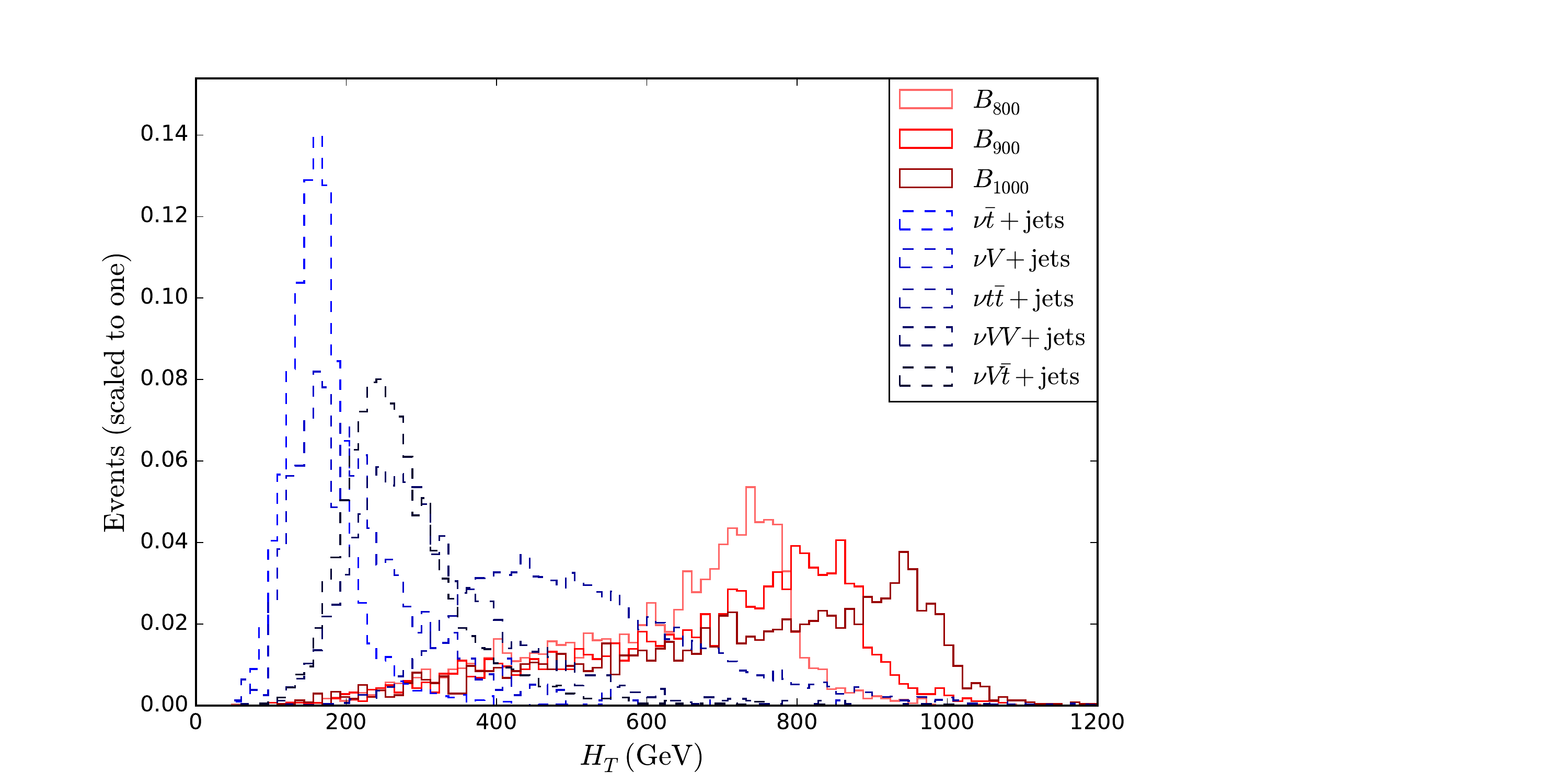}
				\end{minipage}
		\label{fig4d}
	}
	\caption{Normalized distributions of $p^{j}_T$,  $\Delta R^{}_{j_1,j_2}$, $M^{}_{j_1,j_2}$ and $H^{}_T$ for signals and backgrounds
		 at \hspace*{1.65cm} the LHeC ($\sqrt{s}= 1.98$~TeV) with the integrated luminosity of 1000 $\rm fb^{-1}$.}
	\label{fig4}
\end{figure}

We use the characteristics of the signal as a handle to reduce the backgrounds.
Hence, we dipict the normalized distributions of $p^{j}_T$,  $\Delta R^{}_{j_1,j_2}$, $M^{}_{j_1,j_2}$ and $H^{}_T$ for signals and backgrounds in Fig.~\ref{fig4}.
According to the information of these kinematic distributions, we impose the following cuts to get a high statistical significance.
All cuts applied are given in the following list.
\begin{itemize}
	\item Cut 1: The first kinematical selection involves the transverse momentum of jet $p^{j}_T$ , which is shown in Fig.~\ref{fig4a} for signals and backgrounds. 	Since the jets of signal come from the boosted $W$ bosons, they have a larger transverse momentum. Thus, events with $p^{j}_T > 150 $ ~GeV are selected.
	\item Cut 2: The distributions of the transverse momentum $\Delta R^{}_{j_1,j_2}$ of the jets for signals and backgrounds are shown in Fig.~\ref{fig4b}. Based on these normalized distributions, we require the second cut selection is $R^{}_{j_1,j_2} > 2.5$.	
	\item Cut 3: Fig.~\ref{fig4c} and Fig.~\ref{fig4d} show the normalized distributions of $M^{}_{j_1,j_2}$ and $H^{}_T$, where  $M^{}_{j_1,j_2}$ is the invariant mass of two jets and $H^{}_T$ denotes the scalar sum of the transverse momenta of the $b$-tagged jet, the untagged jet and the charged electron. Since the jets of signals all come from the massive VLQ-$B$,
	the $M^{}_{j_1,j_2}$ and $H^{}_T$ peaks of signal and background are separate. Then the cuts $M^{}_{j_1,j_2} > 350$~GeV and $H^{}_T>450$~GeV are imposed.
\end{itemize}

\begin{table*}[htbp]
	\caption{Numbers of  the signal and background events at the LHeC ($\sqrt{s}=1.98$~TeV) with \hspace*{1.65cm}the  integrated luminosity $\mathcal{L}$ = 1000 $\rm fb^{-1}$. Here, we take the coupling parameter $\kappa^{}_B=0.1$  \hspace*{1.65cm} and $R^{}_L = 0.5$.}
	\hspace{15cm}
	\centering
	\begin{tabular}{c|c|c|c|c|c|c|c|c|c}
		\toprule		\hline
		\multicolumn{1}{c|}{}	&\multicolumn{3}{c|}{Signals}	&\multicolumn{6}{c}{\multirow{1}*{Backgrounds}}		\\
		\cline{1-4}	\cline{5-10}
		&$B_{800}$&$B_{900}$&$B_{1000}$
		& $\nu \bar t+$jets & $\nu V+$jets & $\nu t \bar t+$jets & $\nu VV+$jets & $\nu V\bar t+$jets & Total\\	\hline
		Basic cuts	& 242.1 & 110.0 & 53.3 & 546992 & 66100 & 31.8 & 2018.5 & 1001.6 & 616144  \\	\hline
        Cut 1	& 211.6 & 99.4 & 48.5 & 11607 & 11679 & 17.7 & 501.7 & 102.5 & 23908 \\	\hline
        Cut 2	& 205.7 & 97.6 & 47.6 & 7073 & 8125 & 10.8 & 343.6 & 62.9 & 15615 \\\hline
        Cut 3	& 203.9 & 97.3 & 47.4 & 725.5 & 1015.6 & 7.4 & 144.8 & 18.6 & 1911.8 \\		\hline
		\bottomrule
	\end{tabular}
	\label{tab1}
\end{table*}

In order to see whether the signatures of VLQ-$B$ can be detected at the LHeC, we further calculate the statistical significance of signal events:
\begin{equation}
SS = \frac{S}{\sqrt{S+B}},
\label{eq3}
\end{equation}
where $S$ and $B$ denote the numbers of the signal and background events, respectively. We define $SS = 5$ and $ 3 $ as the discovery significance and the possible evidence, respectively.
In Table~\ref{tab1}, we show  the numbers of the signal and background events at the LHeC($\sqrt{s}= 1.98 $TeV) with the integrated luminosity $\mathcal{L} = 1000~ \rm fb^{-1}$.
From the numerical results, we can see that the relevant backgrounds are suppressed effectively, while the signals still have a relatively good efficiency after imposing the above selection cuts. The values of SS can respectively reach about 4.4, 2.2 and 1.0 at the $\mathcal{L}$ = 1000 fb$^{-1}$ for $M_B$=800, 900 and 1000 GeV.

\begin{figure}
	\centering
	\subfigure[]{
		\includegraphics[scale=0.7]{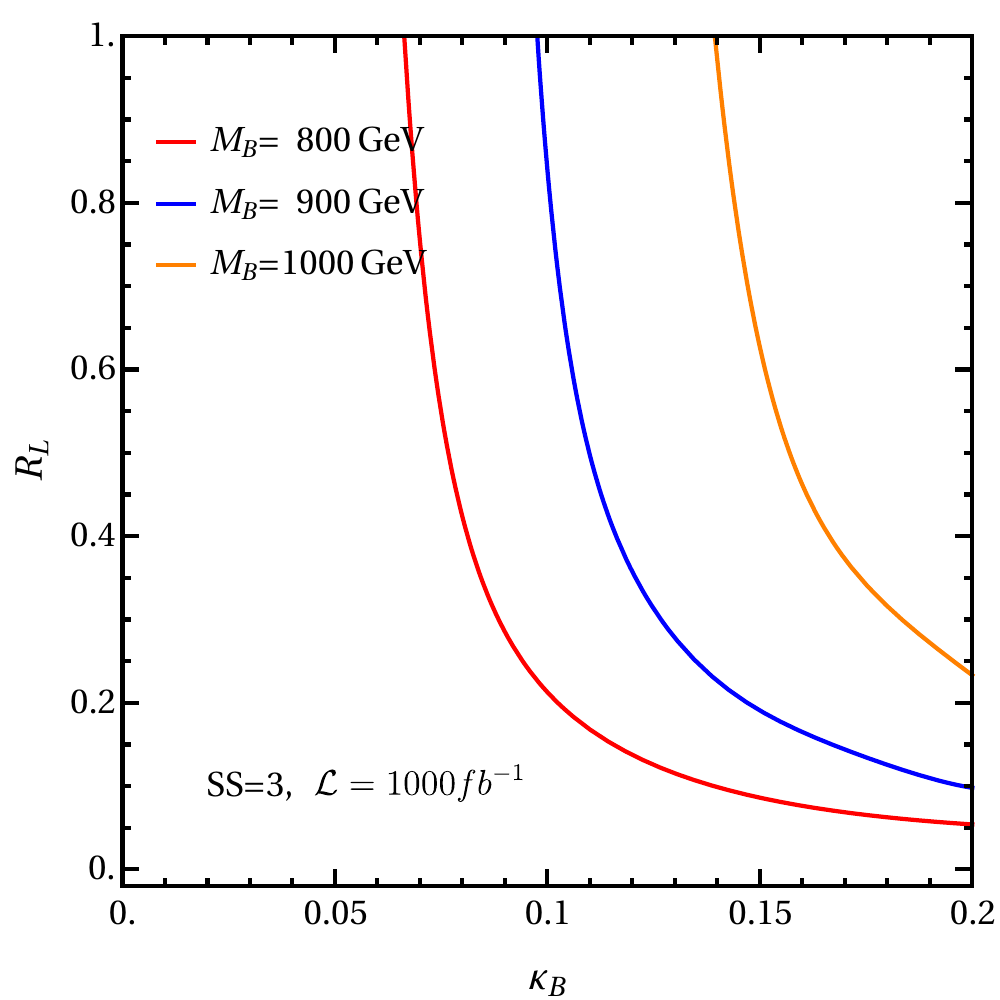}
		\hspace{0.5cm}
	}
	\subfigure[]{
		\includegraphics[scale=0.7]{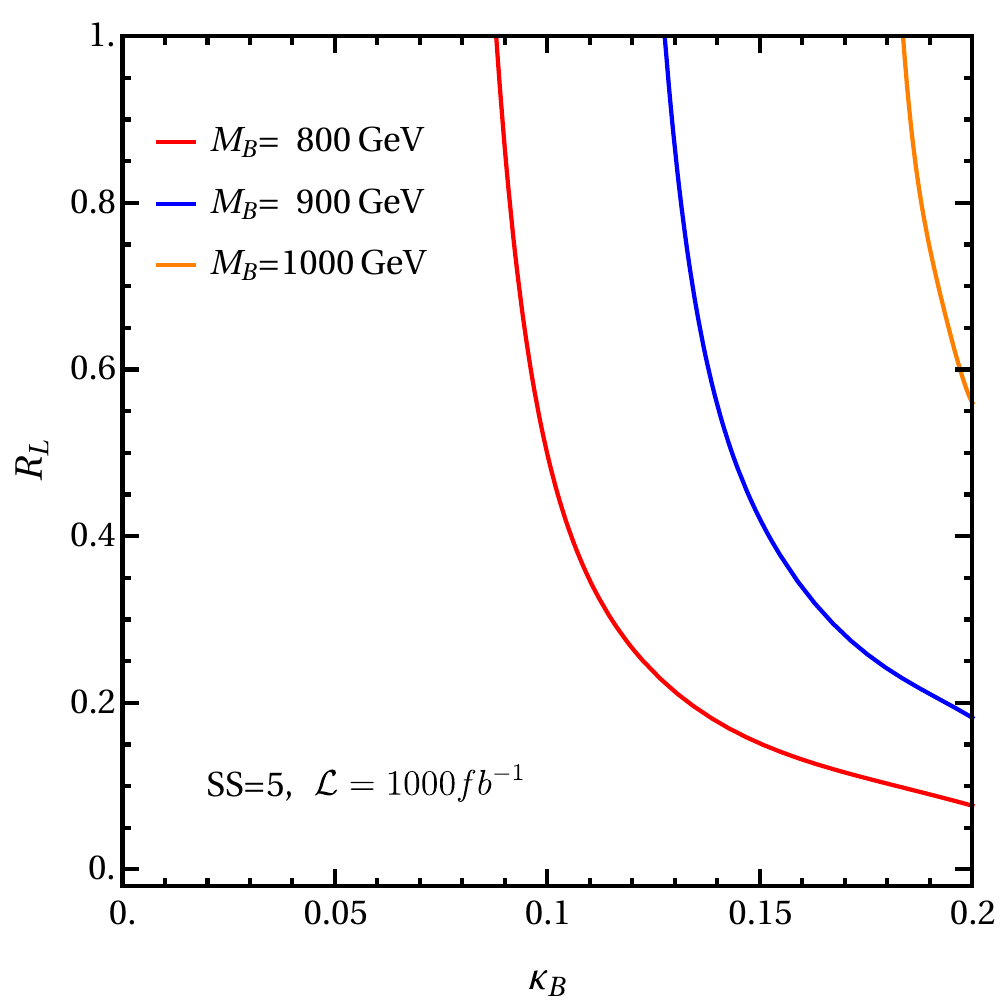}
	}
	\caption{ $3\sigma$(left) and $5\sigma$(right) contour plots in the $R^{}_L - \kappa^{}_B$ plane with three typical VLQ-B \hspace*{1.65cm} masses at the LHeC with the integrated luminosity of 1000 fb$^{-1}$.}
\label{fig5}
\end{figure}

For the purpose of investigating the signal of the singlet VLQ-$B$ more comprehensively, we show the $3\sigma$(left) and $5\sigma$(right) contour plots in the $R^{}_L - \kappa^{}_B$ plane with three typical VLQ-$B$ masses at the LHeC with the integrated luminosity $\mathcal{L} = $1000 fb$^{-1}$  in Fig. \ref{fig5}.
One can see that, the statistical significance can reach 3$\sigma$ when the values of $\kappa^{}_B$  is about 0.076, 0.110 and 0.156 for $R^{}_L = 0.5$ and $M^{}_B$=800, 900 and 1000 GeV respectively.

\subsection*{B. The fully leptonic channel}

Similar with the previous section, we analyze the observation potential and explore the sensitivity of the singlet VLQ-$B$ at the LHeC through the leptonic decay channel:
\begin{equation}
e^- p\rightarrow \nu B(\rightarrow W^-t) \to \nu W^-(\rightarrow l^-_i\bar{\nu}^{}_{i})t(\to W^+b \rightarrow l^+_j\nu^{}_jb) \to l^-_i l^+_j +b+ \slashed{E}_T.
\end{equation}
For this channel, the typical signal is exactly two charged leptons, one $b$ jet, and missing energy.
The main SM backgrounds come from the following processes:
\begin{itemize}
	\item  BKG1: $\nu \bar t$+jets:	$e^- p \to \nu \bar t + {\rm jets} \to \nu W^- \bar b + {\rm jets}\to l^- +\bar b + \slashed{E}_T + {\rm jets}$,
	\item  BKG2: $\nu V$+jets: $e^- p \to \nu V + {\rm jets} \to \left(l^- {\rm or}~ l^-l^+\right) + \slashed{E}_T + {\rm jets}$,
	\item  BKG3: $\nu V V$+jets: $e^- p \to \nu V V + {\rm jets} \to \left(2l,3l~{\rm or}~4l\right) + \slashed{E}_T + {\rm jets} $,
	\item  BKG4: $\nu V \bar t$+jets: $e^- p \to \nu Z \bar t + {\rm jets} \to \nu l^-l^+ W^- \bar b + {\rm jets} \to 2l^-+ l^+ + \bar b + \slashed{E}_T + {\rm jets}, $
\end{itemize}
where the light jet can be  faked as $b$-jet and the leptons may escape the detector.

In our simulation, we apply the following basic cuts on the signal and background events firstly
$$ p^{l}_{T}> 10~{\rm GeV} \;,\quad  \vert\eta^{l} \vert<2.5 \;,\quad  p^{j}_{T} > 20~{\rm GeV} \;,\quad
\vert\eta^{j} \vert<5 \;,\quad \Delta R(x,y) > 0.4 \;,\quad x,y = l, j.$$

\begin{figure}[http]
	\centering
	\subfigure[]{
		\begin{minipage}[t]{0.48\textwidth}
			\includegraphics[height=6cm, width=11cm]{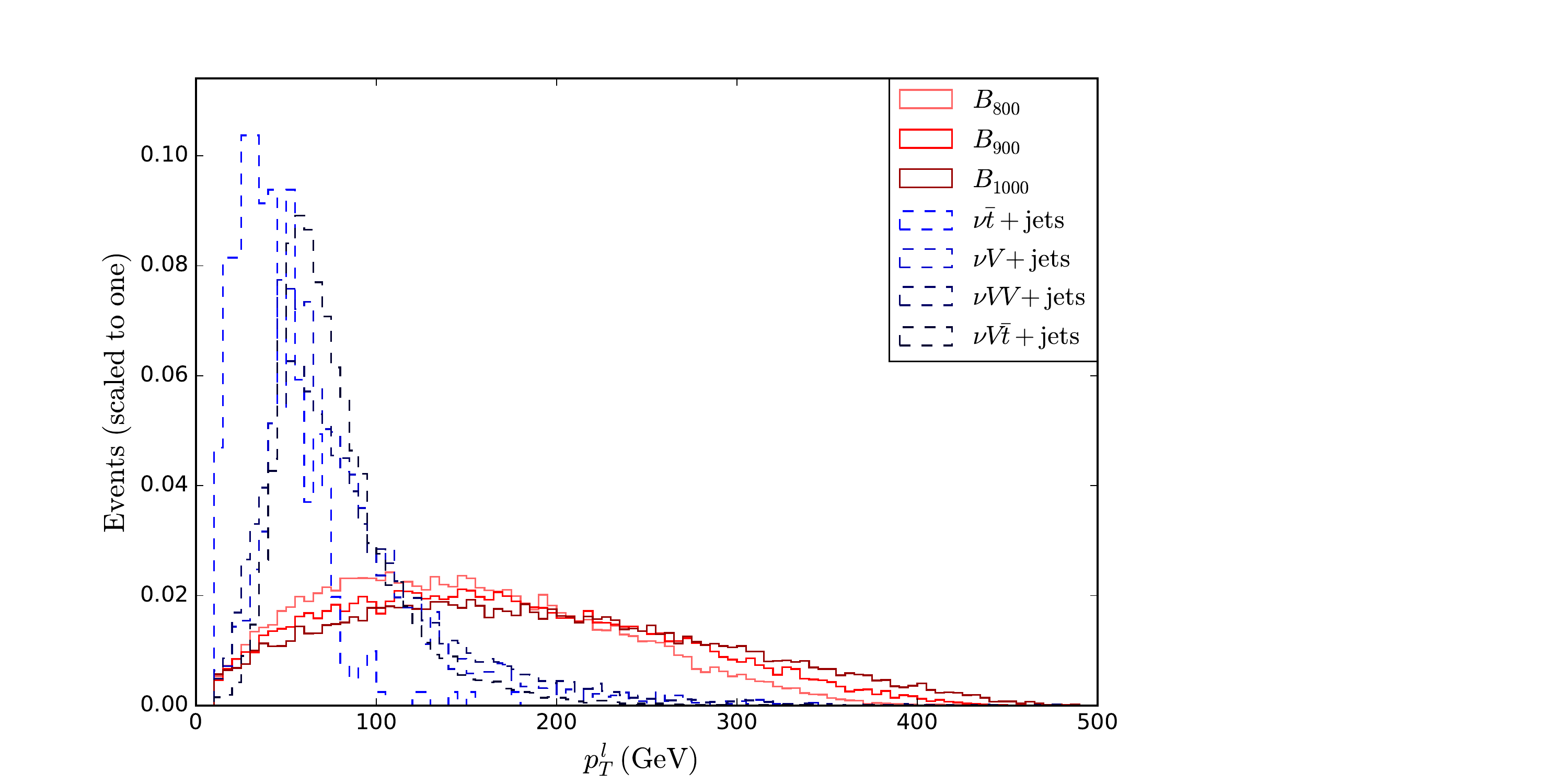}
		\end{minipage}
		\label{fig6a}}
	\centering
	\subfigure[]{
		\begin{minipage}[t]{0.48\textwidth}
			\includegraphics[height=6cm, width=11cm]{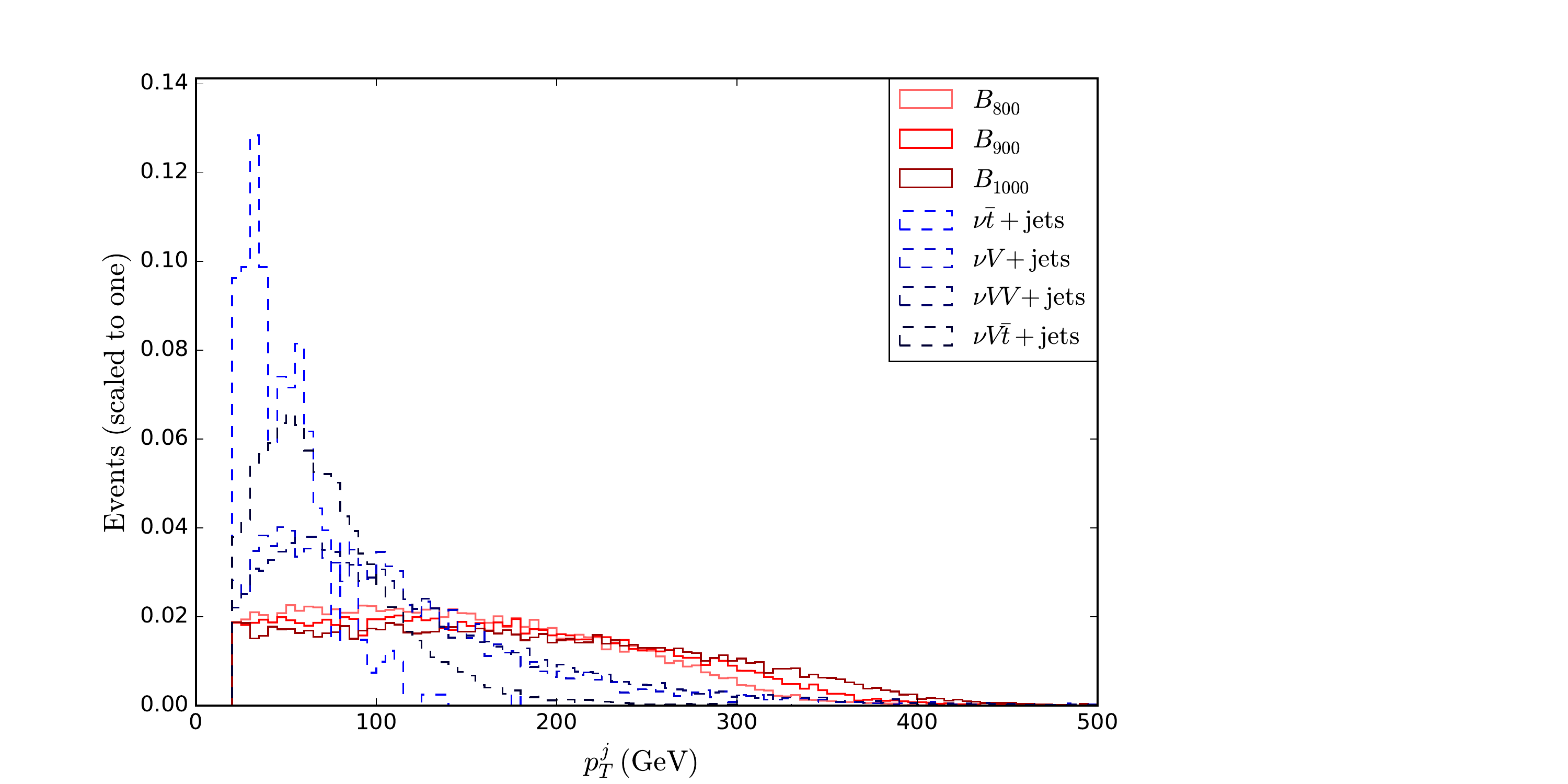}
		\end{minipage}
		\label{fig6b}
	}		
	\centering
	\subfigure[]{
		\begin{minipage}[b]{0.48\textwidth}
			\includegraphics[height=6cm, width=11cm]{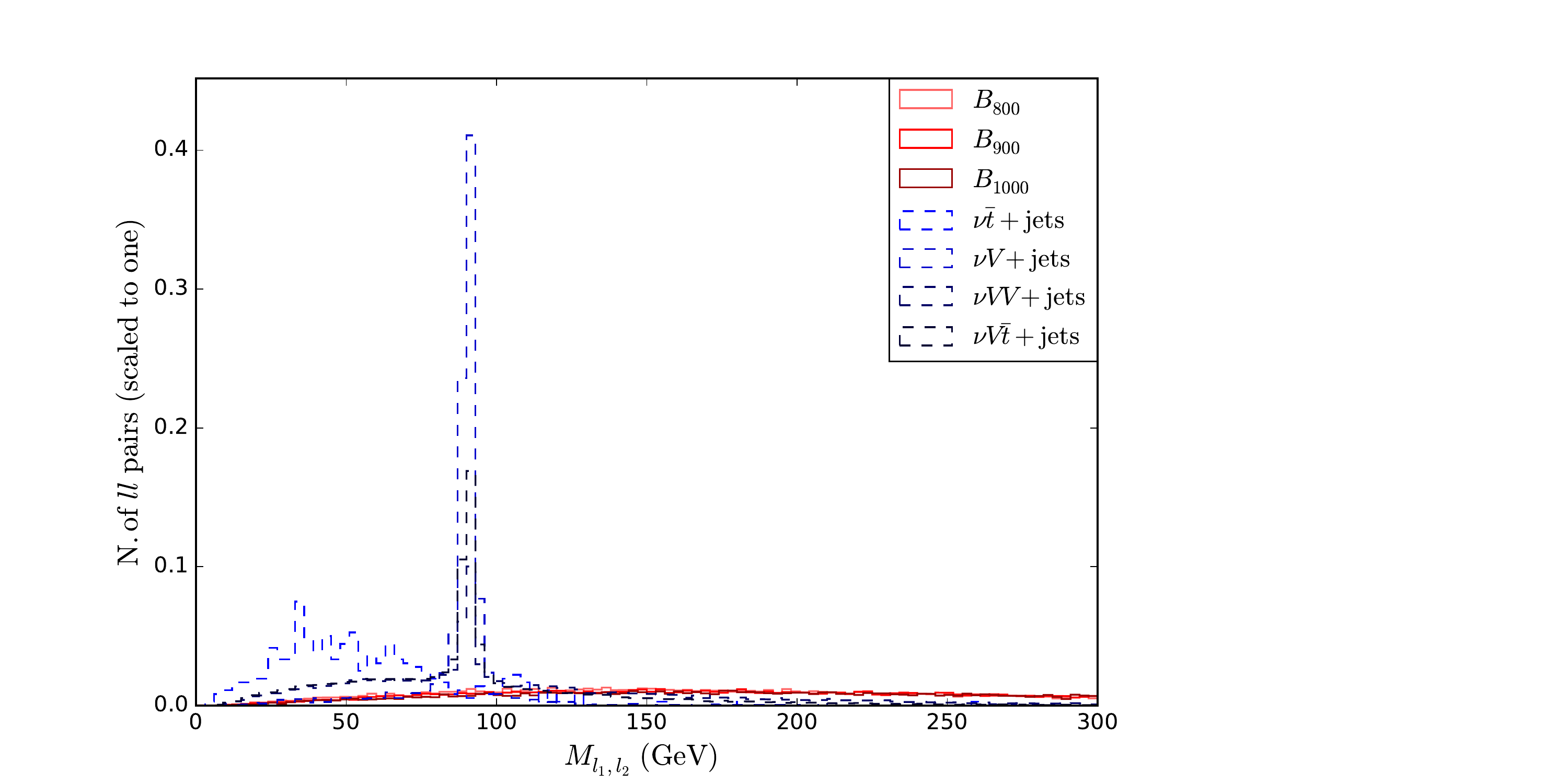}
		\end{minipage}
		\label{fig6c}
	}
	\centering
\subfigure[]{
	\begin{minipage}[b]{0.48\textwidth}
		\includegraphics[height=6cm, width=11cm]{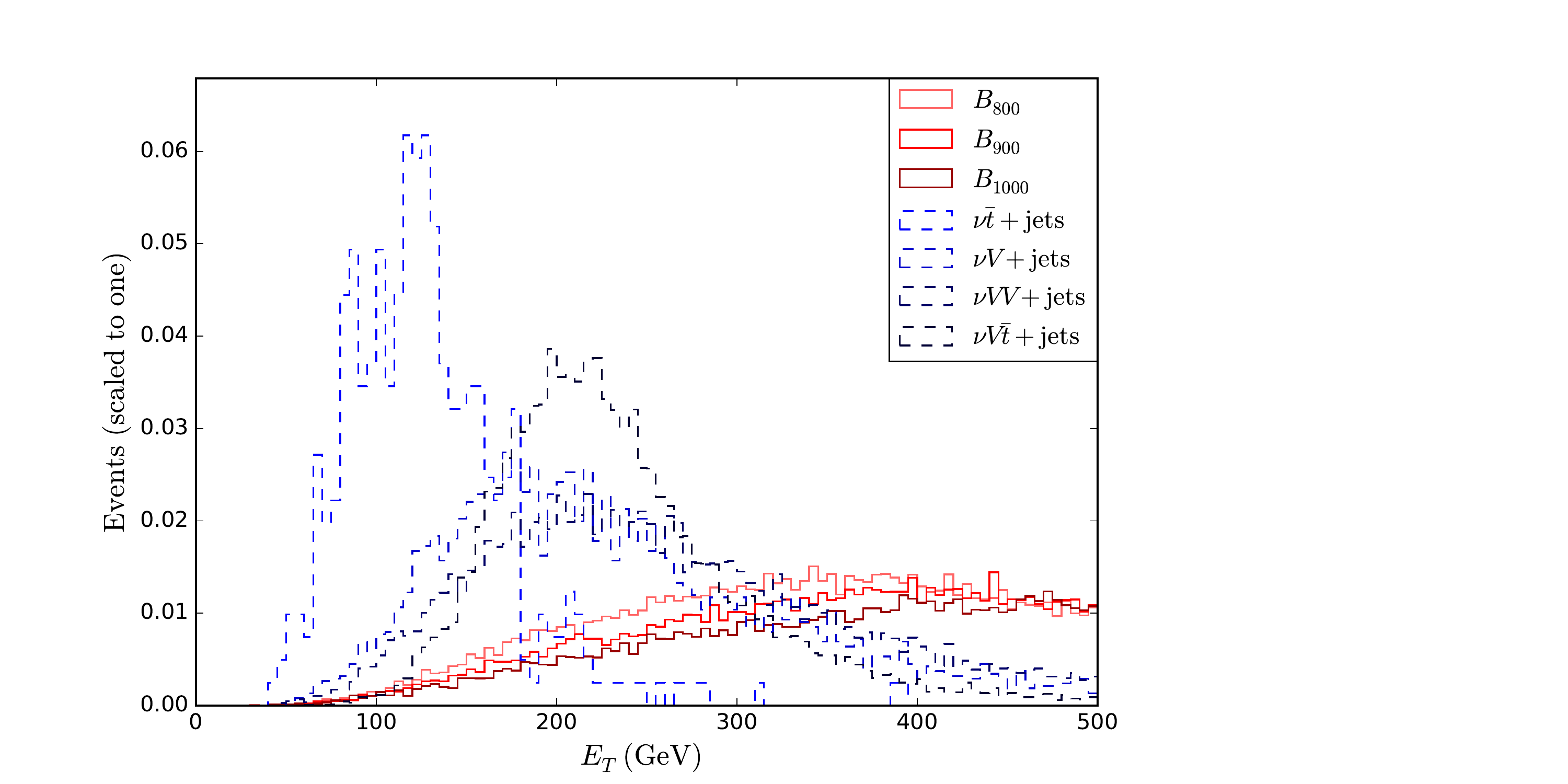}
	\end{minipage}
	\label{fig6d}
}
	\caption{Normalized distributions of $p^{l}_{T}$, $p^j_T$, $M^{}_{j_1,j_2}$ and $E^{}_T$ for signals and backgrounds at the \hspace*{1.65cm}  LHeC with the integrated luminosity of 1000 $\rm fb^{-1}$.}
	\label{fig6}
\end{figure}
\begin{table*}[htbp]
	\caption{Same as Table 1, but for the fully leptonic channel.}
	\hspace{15cm}
	\centering
	\begin{tabular}{c|c|c|c|c|c|c|c|c}
		\toprule	\hline
		\multicolumn{1}{c|}{}	&\multicolumn{3}{c|}{Signals}		&\multicolumn{5}{c}{\multirow{1}*{Backgrounds}}		\\
		\cline{1-4}		\cline{5-9}
		& $B_{800}$ & $B_{900}$ & $B_{1000}$ & $\nu \bar t$+jets & $\nu V$+jets & $\nu VV$+jets & $\nu V \bar t$+jets & Total \\		\hline 	
        Basic cuts	& 171.5 & 99.0 & 53.2 & 23230 & 116730 & 1374.8 & 108.9 & 141444  \\\hline
        Cut 1	& 112.9 & 69.7 & 38.7 & 229.4 & 42667 & 521.1 & 25.6 & 43443 \\	\hline
        Cut 2	& 92.7 & 58.6 & 33.1 & 114.7 & 2391.1 & 347.7 & 14.8 & 2868.3 \\\hline
        Cut 3	& 81.3 & 52.9 & 30.7 & 57.4 & 1304.2 & 244.9 & 8.1 & 1614.6 \\	\hline
		\bottomrule
	\end{tabular}
	\label{tab2}
\end{table*}

In order to get some hints of further cuts for reducing the backgrounds,  we analysis the normalized distributions of $p^{l}_{T}$, $p^j_T$, $M^{}_{j_1,j_2}$ and $E^{}_T$ for the signals and backgrounds as shown in Fig.~\ref{fig6}.
Then, to get  high statistical significance, a set of further cuts are given as followings.

\begin{itemize}
	\item Cut 1: The normalized distributions of transverse momenta of leptons and jets for signals and backgrounds are shown in Fig. ~\ref{fig6a} and Fig. ~\ref{fig6b} , we can see that the transverse momenta of signal events are distributed mostly at large $p^{l,j}_T$ values, which are different from the distributions of background events.
	Then we require $p^{l}_T > 70$ GeV and $p^{j}_T > 70$ GeV   to enhance the signal significance.
	\item Cut 2: In Fig. ~\ref{fig6c}, the background events have sharp peaks in the distributions of the invariant mass of lepton pairs.
	So the cut $M^{}_{j_1,j_2} > 100$ GeV is applied.
	\item Cut3: Fig.~\ref{fig6d} show the normalized distributions of $E^{}_T$ which denotes the total transverse energy of leptons and jets.
	From these distributions, we can efficiently reduce the backgrounds by the cut:~$ E^{}_T > 300$ GeV.
\end{itemize}
\begin{figure}
	\centering
	\subfigure[]{
		\includegraphics[scale=0.7]{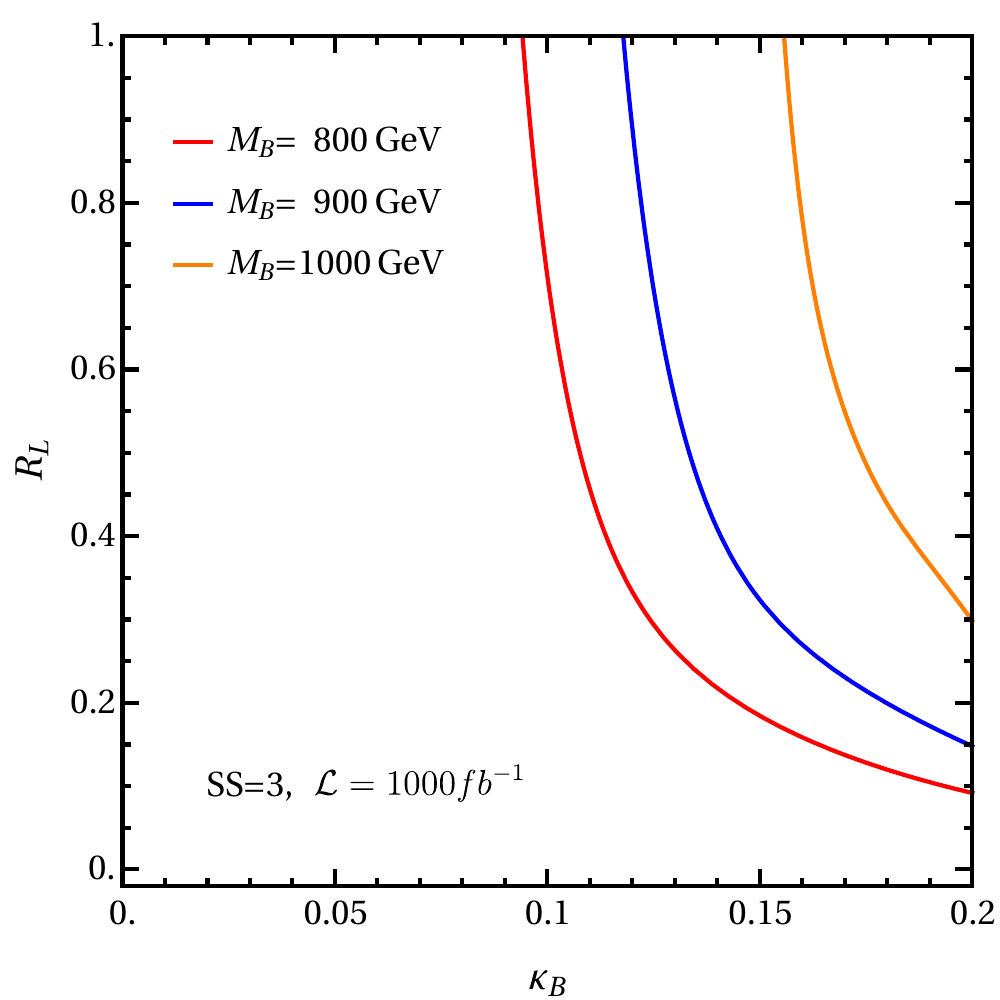}
		\hspace{0.5cm}
	}
	\subfigure[]{
		\includegraphics[scale=0.7]{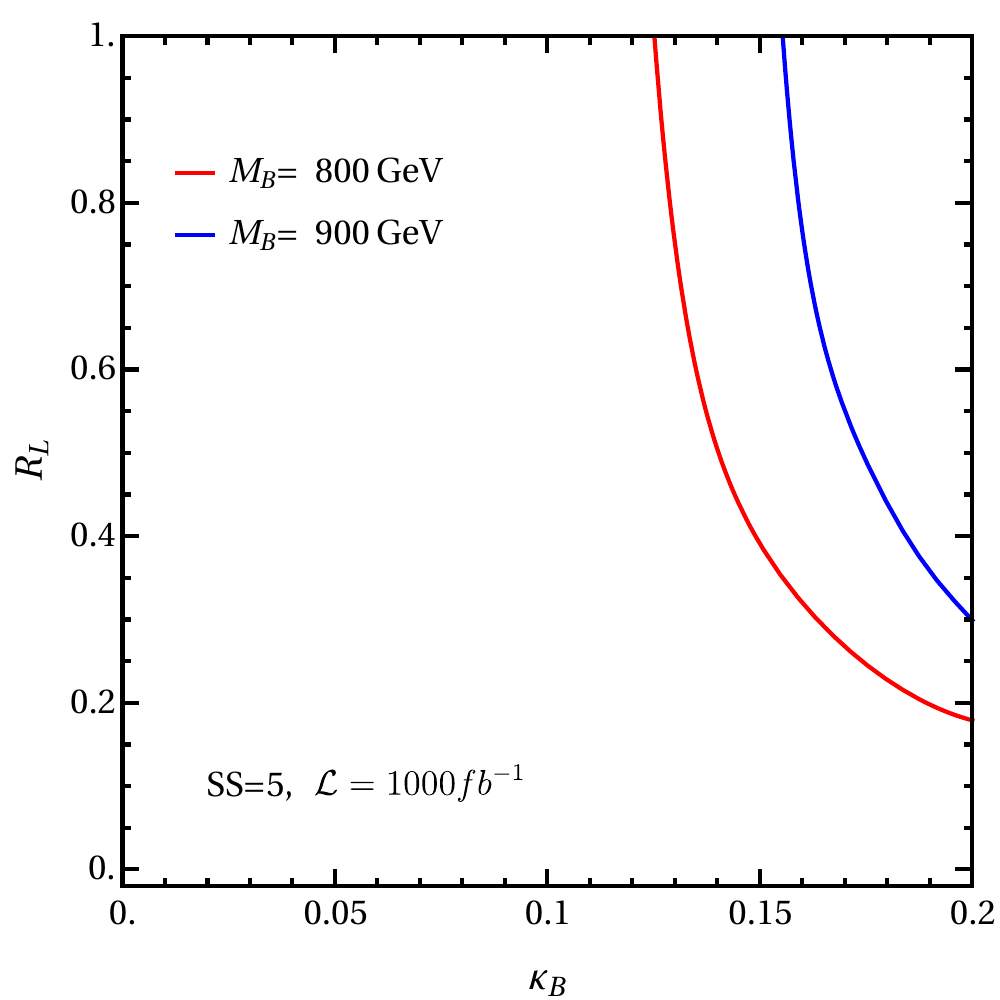}
	}
	\caption{Same as Fig.5, but for the fully leptonic channel.}
	\label{fig7}
\end{figure}

We summarize the numbers of the signal and background events at the LHeC ($\sqrt{s}= 1.98$~TeV) with the integrated luminosity $\mathcal{L}$ = 1000 $\rm fb^{-1}$ in Table~\ref{tab2}.
The values of SS can respectively reach about 2.0, 1.3 and 0.8 at the $\mathcal{L}$ = 1000 fb$^{-1}$ for $M_B$=800, 900 and 1000 GeV.
And the $3\sigma$(left) and $5\sigma$(right) contour plots in the $R^{}_L - \kappa^{}_B$ plane are shown in Fig.~\ref{fig7}.
We can see that, for $R^{}_L = 0.5$ and $M^{}_B$=800, 900 and 1000 GeV, SS can reach  3$\sigma$ when the values of $\kappa^{}_B$ is about 0.107, 0.132 and 0.173. Compared to the fully hadronic channel, the fully leptonic channel is not a good process to test the singlet VLQ-$B$ at the LHeC.

\subsection*{C. The semileptonic channel }

 Now, we investigate the observability of the singlet VLQ-$B$ at the LHeC  through the semileptonic decay channel
\begin{equation}
e^- p\rightarrow \nu B(\rightarrow W^-t) \to \nu W^-(\rightarrow l^-_i\bar{\nu}^{}_{i})t(\to W^+b \rightarrow j j b) \to l^- + 2j + b+ \slashed{E}_T.
\end{equation}
For this channel, the typical signal is exactly one charged lepton, one $b$-jet, two jets (which coming from the top quark decay) and missing energy.

The dominant SM backgrounds come from the following processes:
\begin{itemize}
	\item  BKG1: $\nu \bar t$+jets:	$e^- p \to \nu \bar t + {\rm jets} \to \nu W^- \bar b + {\rm jets}\to l^- +\bar b + \slashed{E}_T + {\rm jets}$,
	\item  BKG2: $\nu V$+jets: $e^- p \to \nu V + {\rm jets} \to \left(l^- {\rm or}~ l^-l^+\right) + \slashed{E}_T + {\rm jets}$,
	\item  BKG3: $\nu V V$+jets: $e^- p \to \nu V V + {\rm jets} \to \left(l^- {\rm or}~ l^-l^+\right) + \slashed{E}_T + {\rm jets} $,
	\item  BKG4: $\nu V \bar t$+jets: $e^- p \to \nu Z \bar t + {\rm jets} \to \nu Z W^- \bar b + {\rm jets} \to 2l^- +l^+ + \bar b + \slashed{E}_T + {\rm jets}, $
\end{itemize}
where one light jet might be faked as $b$-jet and the leptons may escape the detector.

\begin{figure}[http]
	\subfigure[]{
		\begin{minipage}[t]{0.5\textwidth}
			\includegraphics[height=6cm, width=11cm]{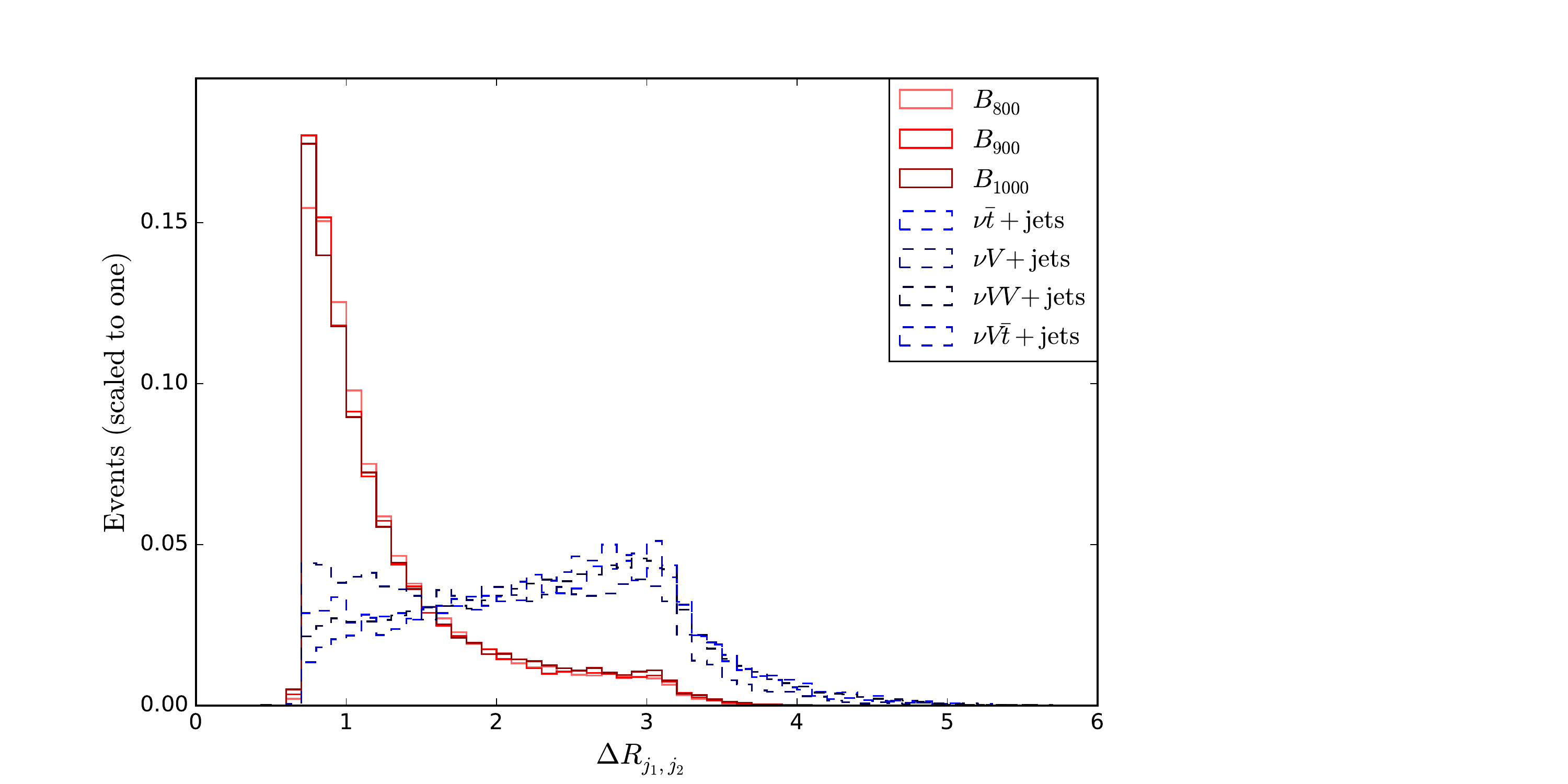}
		\end{minipage}
		\label{fig8a}}
	\subfigure[]{
		\includegraphics[height=6cm, width=11cm]{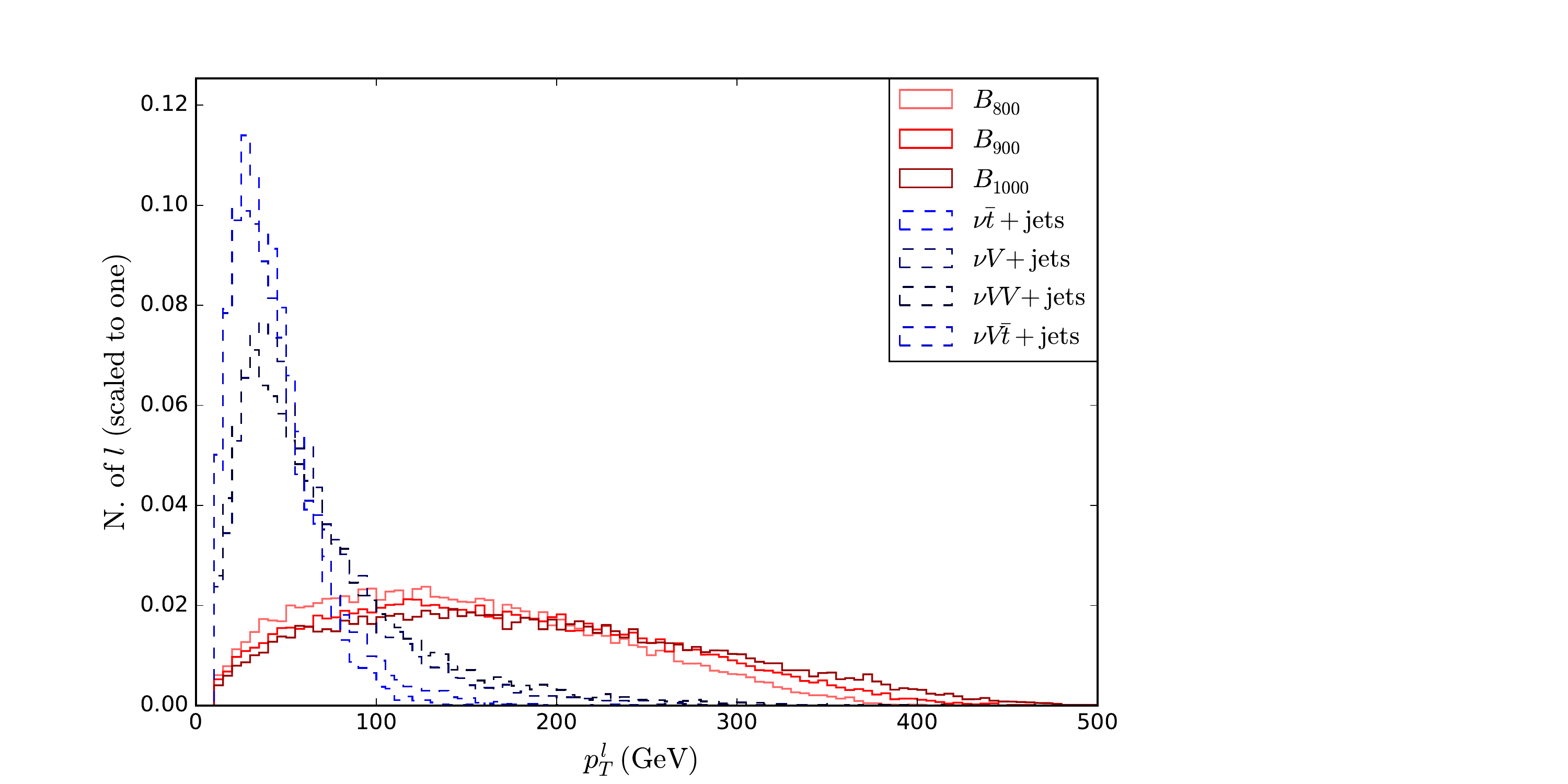}
		\label{fig8b}
	}		
	
	\subfigure[]{
		\begin{minipage}[t]{0.5\textwidth}	
			\includegraphics[height=6cm, width=11cm]{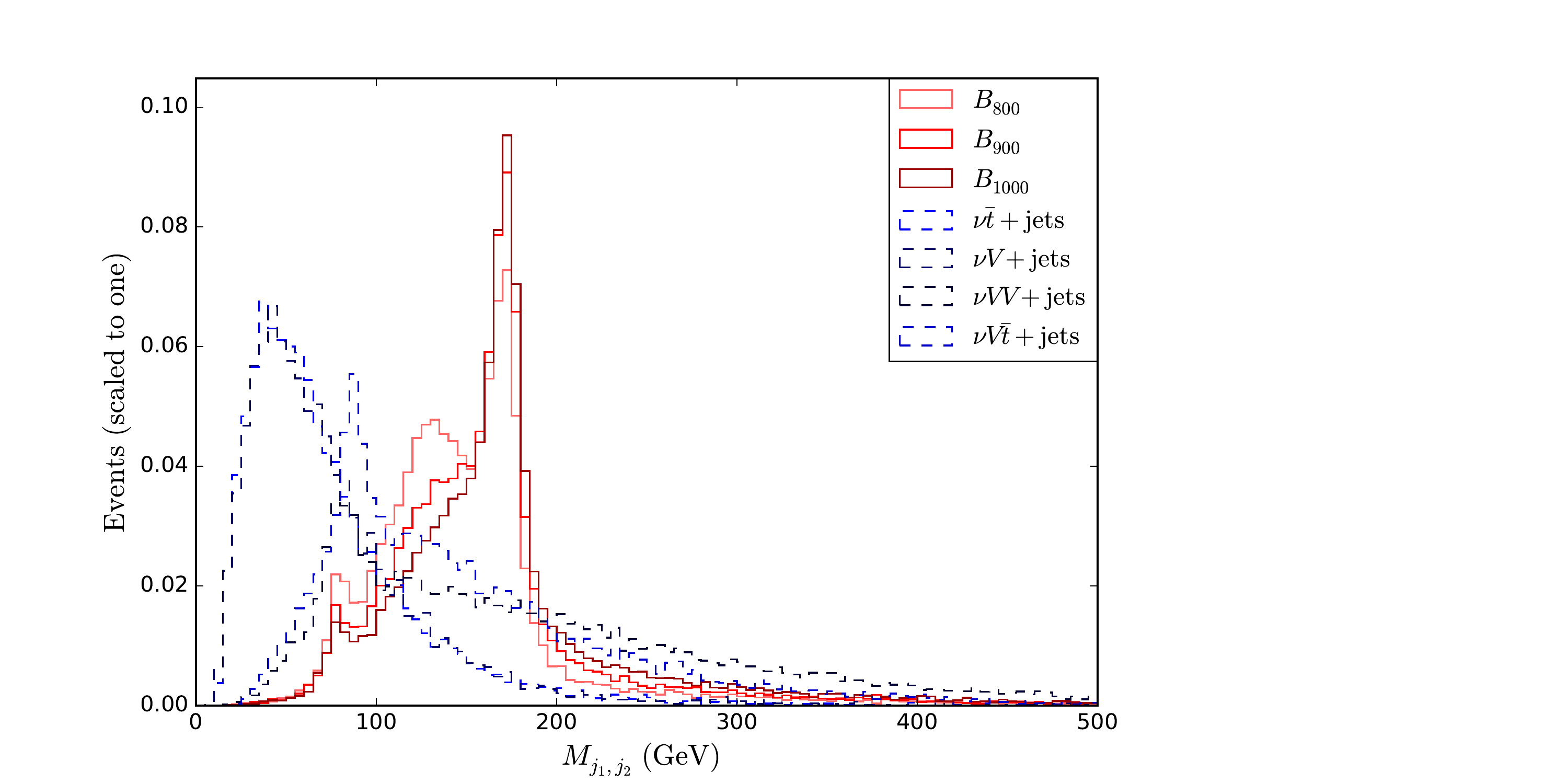}	
		\end{minipage}	
		\label{fig8c}
	}
	\subfigure[]{
		\begin{minipage}[t]{0.5\textwidth}	
			\includegraphics[height=6cm, width=11cm]{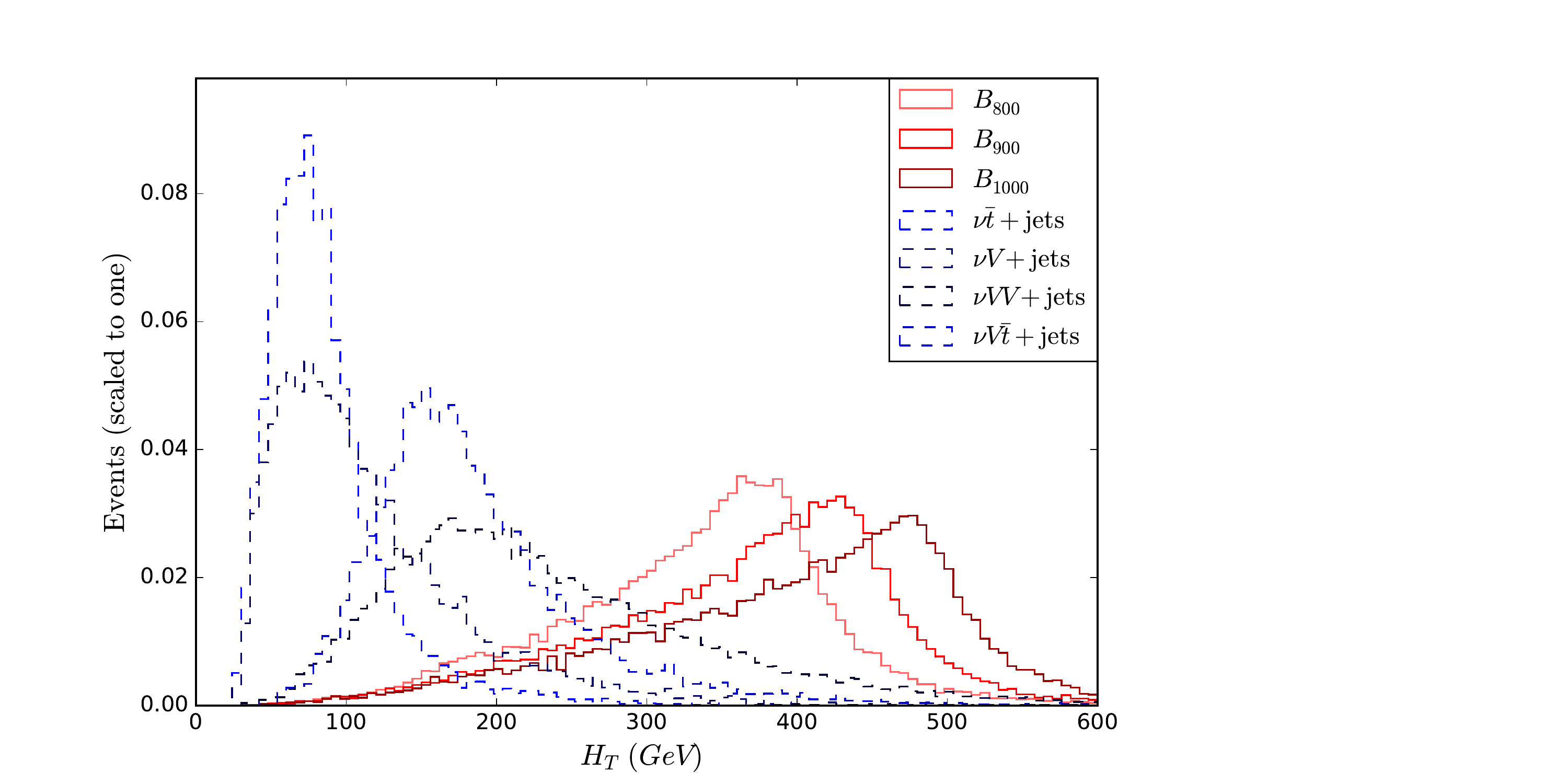}	
		\end{minipage}	
		\label{fig8d}
	}
	\caption{Normalized distributions of  $\Delta R^{}_{j_1,j_2}$,  $p^l_T$, $M^{}_{j_1,j_2}$ and $H_T$ for signals and backgrounds at \hspace*{1.65cm} the LHeC  with the integrated luminosity of 1000 $\rm fb^{-1}$.}
	\label{fig8}
\end{figure}

We apply the following basic cuts on the signal and background events in our simulation:
\begin{eqnarray}
p^{l}_{T}> 10~{\rm GeV} \;,\quad \vert\eta^{l}\vert<2.5 \;,\quad p^{j}_{T}> 20~{\rm GeV} \;,\quad \vert\eta^{j} \vert<5 \;,\quad \Delta R(x,y)>0.4,\ ~~x,y = l, j
\end{eqnarray}
Further, we apply some general preselections as following.

To carry out the cut-based analysis, we discuss the normalized distributions of $\Delta R^{}_{j_1,j_2}$,  $p^l_T$, $M^{}_{j_1,j_2}$ and $H_T$  for signals and backgrounds at the LHeC with the integrated luminosity of 1000 $\rm fb^{-1}$ shown in Fig.~\ref{fig8}.
All cuts are applied one after the other in the order given in the following list.
\begin{itemize}
\item Cut 1 : In Fig.~\ref{fig8a}, we show the normalized distributions of the particle separation of jets $\Delta R_{j_1,j_2}$ for signals and backgrounds.
Since most jets of signals come from boosted object, they have much narrower separation. Based on the normalized distributions, we impose the first cut to get a high significance: $ \Delta R_{j_1,j_2} < 1.5$.
\item Cut 2 : Fig.~\ref{fig8b} is the normalized distributions of the transverse momenta of leptons. Since the leptons of backgrounds most come from the decay of static $W$ or $Z$ bosons,
the peaks local in half the mass of the $W$ or $Z$ boson. Thus cut $p^l_T > 80$ GeV can reduce the backgrounds effectively.
\item Cut 3: In Fig.\ref{fig8c}, the normalized distribution of $M^{}_{j_1,j_2}$ is given. The peaks of signals are larger than those of backgrounds. Then the cut with $M^{}_{j_1,j_2} > 100$ GeV is selected.
\item Cut 4:  The Fig.\ref{fig8d} show the total hadronic energy $H^{}_T$. Since the massive VLQ-$B$, the signal events should have higher hadronic energy than those of background events. Finally, we keep the events with $H^{}_T > 300 $ GeV.
\end{itemize}

\begin{table*}[htbp]
	\caption{Same as Table 1 but for the semileptonic channel.}
	\hspace{15cm}
	\centering
	\begin{tabular}{c|c|c|c|c|c|c|c|c}
		\toprule	\hline
		\multicolumn{1}{c|}{}	&\multicolumn{3}{c|}{Signals}	&\multicolumn{5}{c}{\multirow{1}*{Backgrounds}}		\\
		\cline{1-4}	\cline{5-9}
		& $B_{800}$ & $B_{900}$ & $B_{1000}$ & $\nu \bar t$+jets & $\nu V$+jets & $\nu VV$+jets & $\nu V \bar t$+jets & Total \\\hline
        Basic cuts 	& 629.1 & 369.6 & 199.0 & 497564 & 159070 & 9306.5 & 828.8 & 666770 \\\hline
        Cut 1 & 472.9 & 278.2 & 147.2 & 114242 & 50103 & 1950.7 & 144.6 & 166441 \\\hline
        Cut 2 & 378.7 & 232.3 & 125.5 & 7539.8 & 13289 & 706.4 & 20.0 & 21555 \\ \hline
       Cut 3 & 332.5 & 211.5 & 116.9 & 241.3 & 667.5 & 308.2 & 3.1 & 1220.1 \\ \hline
       Cut 4 & 276.2 & 187.6 & 106.4 & 0 & 283.2 & 148.5 & 0.8 	& 432.5 \\
       \hline
		\bottomrule
	\end{tabular}
	\label{tab3}
\end{table*}

\begin{figure}
	\centering
	\subfigure[]{
		\includegraphics[scale=0.7]{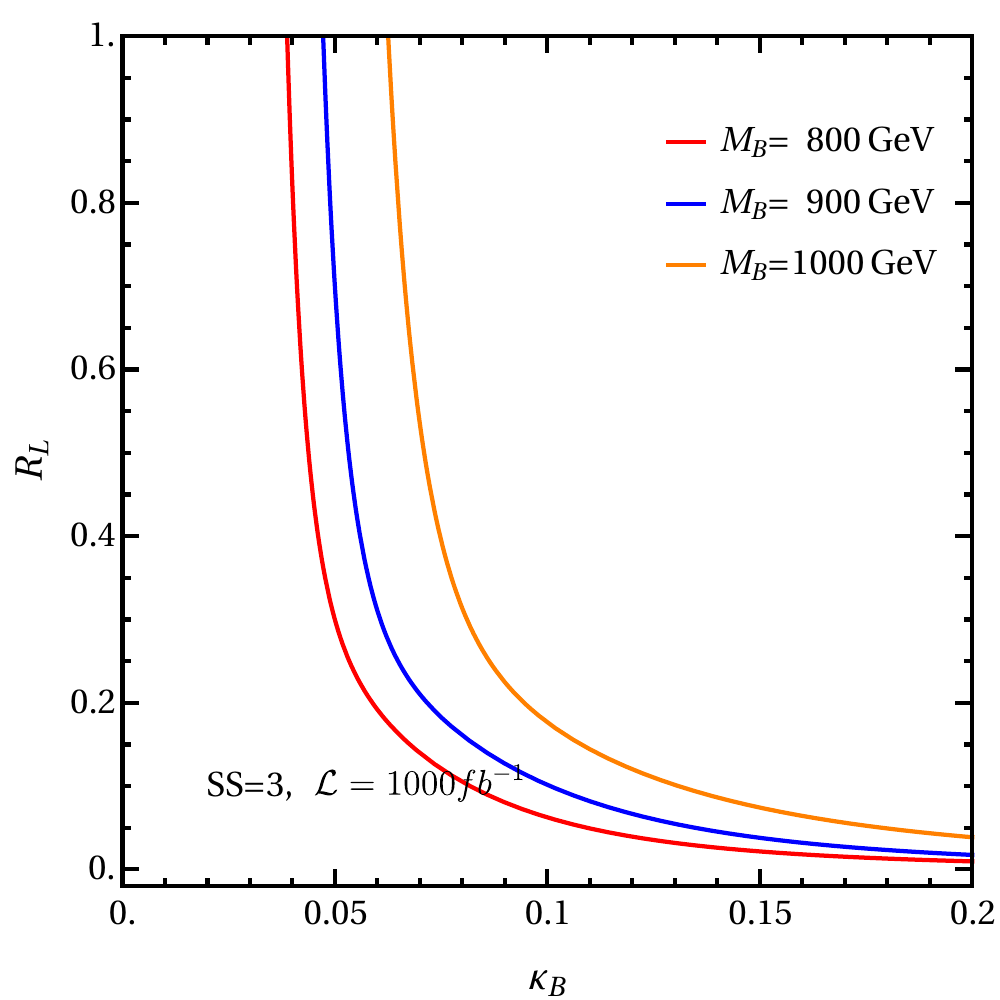}
		\hspace{0.5cm}
	}
	\subfigure[]{
		\includegraphics[scale=0.7]{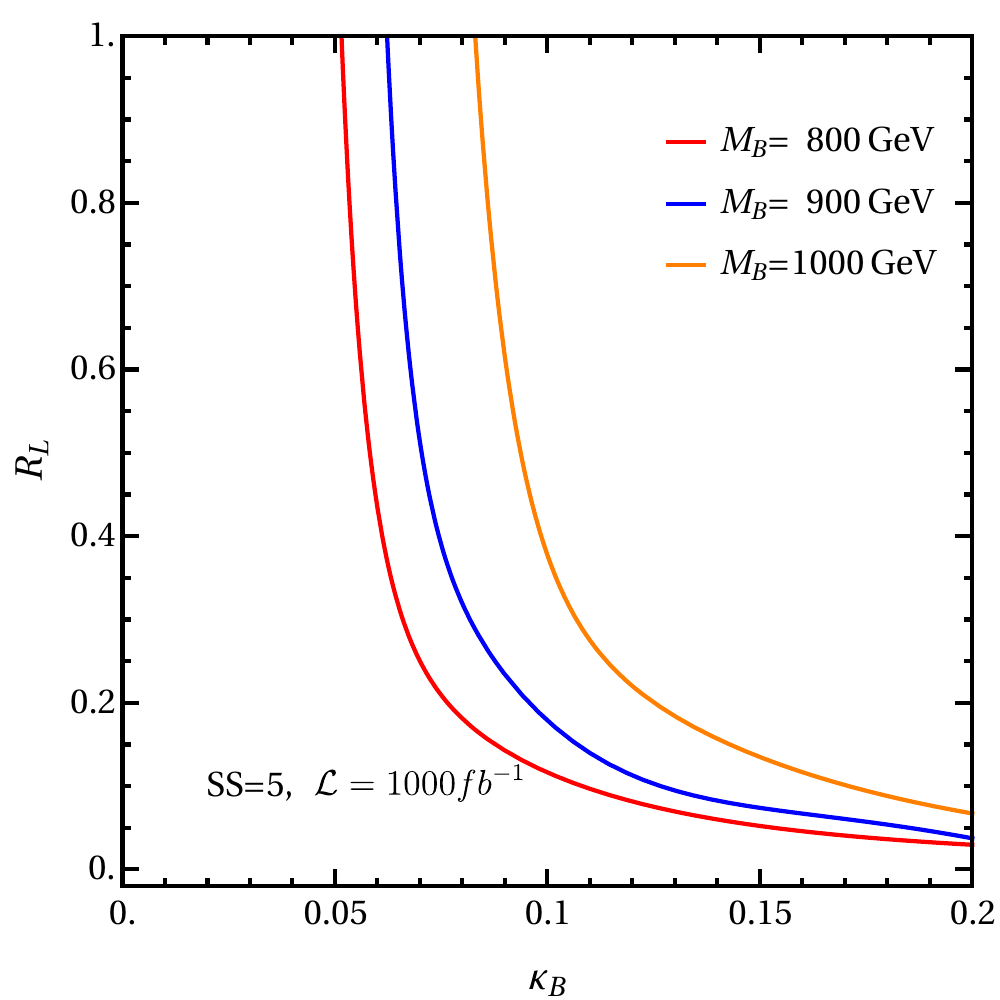}
	}
	\caption{Same as Fig.5, but for the semileptonic channel.}
	\label{fig9}
\end{figure}

In Table~\ref{tab3}, we show the numbers of the signal and background events at the LHeC ($\sqrt{s}= 1.98$~TeV) with the integrated luminosity $\mathcal{L}$ = 1000 $\rm fb^{-1}$.
After imposing the above selection cuts,  the backgrounds are suppressed efficiently.
The values of SS can respectively reach about 10.4, 7.5 and 4.6 at the $\mathcal{L}$ = 1000 fb$^{-1}$ for $M_B$=800, 900 and 1000 GeV.
Fig.~\ref{fig9} shows the $3\sigma$(left) and $5\sigma$(right) contour plots in the $R^{}_L - \kappa^{}_B$ plane with three typical VLQ-$B$ mass at the LHeC with $\mathcal{L}$ = 1000 fb$^{-1}$.
For $R^{}_L = 0.5$ and $M^{}_B$=800, 900 and 1000 GeV, SS reach 3$\sigma$ (5$\sigma$) while the values of $\kappa^{}_B$ achieve about 0.043(0.058), 0.053(0.070) and 0.071(0.093) separately.

From above discussions, we can see that, for single production of  VLQ-$B$ at the LHeC, it is possible to detect its signal via the fully hadronic, the fully leptonic and the semileptonic final states.
However,  the VLQ-$B$ which is the $SU(2)$ singlet with electric charge $-1/3$, can be more easy detected via the semileptonic decay channel at the LHeC.

\section*{V. Conclusions}

In this paper, we study the discovery potential of the single production of VLQ-$B$ at the LHeC, through three types of the  characteristic signals, which come from the fully hadronic, the fully leptonic and the semileptonic decay channels. We focus our attention on the $SU(2)$ singlet VLQ-$B$ with electric charge $-1/3$ in a model-independent fashion.
We investigate the observability of the VLQ-$B$ signal through these three decay channels at the LHeC with the integrated luminosity $\mathcal{L}$ = 1000 $\rm fb^{-1}$.
In our numerical calculation, we obtain the 3$\sigma$ possible evidence region as well as the 5$\sigma$ discovery region, which  are respectively presented in terms of parameter space regions  for three typical masses (800 GeV, 900 GeV, 1000 GeV).
For $R^{}_L = 0.5$ and $M^{}_B$=800, 900 and 1000 GeV, the values of SS reach 3$\sigma$ when $\kappa^{}_B$ is about 0.076, 0.110 and 0.156  in the fully hadronic channel, 0.107, 0.132 and 0.173 in the fully leptonic channel, and 0.043, 0.053 and 0.071 in the semileptonic channel.
Thus, the possible signatures of the $SU(2)$ singlet VLQ-$B$ with electric charge $-1/3$ is easier detected via the process $e^- p \to \nu B(\rightarrow W^{-}t)$ in the semileptonic channel than other decay channels at the LHeC. We expect our analysis can provide a complementary candidate to pursue searching for the singlet VLQ-$B$ at the LHeC.

\section*{ACKNOWLEDGMENT}
This work was partially supported by the National Natural Science Foundation of China under Grants No. 11875157 and Grant No. 11847303.

\end{document}